\documentstyle[prb,aps,twocolumn,epsf,floats]{revtex} 
\begin{document} 
\draft 
\twocolumn[\hsize\textwidth\columnwidth\hsize\csname @twocolumnfalse\endcsname 
\title{Theory of Orbital State and Spin Interactions 
in Ferromagnetic Titanates} 
\author{Giniyat~Khaliullin} 
\address{Max-Planck-Institut f\"ur Festk\"orperforschung,  
Heisenbergstrasse 1, D-70569 Stuttgart, Germany}  
\author{Satoshi~Okamoto\cite{columbia}}
\address{The Institute of Physical and Chemical Research (RIKEN), 
Saitama 351-0198, Japan} 
\date{\today} 
\maketitle 
\begin{abstract} 
A spin-orbital superexchange Hamiltonian in a Mott insulator with  
$t_{2g}$ orbital degeneracy is investigated.  
More specifically, we focus on a spin ferromagnetic state of the model  
and study a collective behavior of orbital angular momentum.  
Orbital order in the model occurs in a nontrivial way
-- it is stabilized exclusively by quantum effects 
through the order-from-disorder mechanism.  
Several energetically equivalent orbital orderings are identified. 
Some of them are specified by a quadrupole ordering and 
have no unquenched angular momentum at low energy.  
Other states correspond to a noncollinear ordering of the  
orbital angular momentum and show the 
magnetic Bragg peaks at specific positions.  
Order parameters are unusually small because of strong quantum fluctuations.  
Orbital contribution to the resonant x-ray scattering is discussed.   
The dynamical magnetic structure factor in different 
ordered states is calculated.  
Predictions made should help 
to observe elementary excitations of orbitals and also  
to identify the type of the orbital order in ferromagnetic titanates. 
Including further a relativistic spin-orbital coupling,  
we derive an effective low-energy spin Hamiltonian 
and calculate a spin-wave spectrum,  
which is in good agreement with recent experimental observations in YTiO$_3$. 
\end{abstract} 
] 
\narrowtext 
\noindent 
 
\section{Introduction} 
Many transition-metal oxides fall into the category of Mott 
insulators,\cite{IMA98}  
in which the large degeneracy of atomic states remain unquenched down to low
energies.   
Of particular importance here is the role being played by orbital degeneracy  
inherent to perovskite lattices.  
An additional degeneracy of low-energy states and the extreme sensitivity of  
the chemical bonds to the spatial orientation of orbitals lead to  
frustrating interactions and a variety of competing phases  
that are tunable by moderate external fields.\cite{TOK00} 
 
As ``orbital physics'' has started to become an essential  
ingredient of the physics of transition-metal oxides,  
more efforts are necessary to develop quantum many-body theory of  
coupled spin-orbital systems in order to understand 
specific features of the orderings and fluctuations in these models.  
Earlier work  
has emphasized a ``classical part'' of the problem, 
focusing mainly on the strong interplay  
between classical spin and orbital configurations.  
It is implicitly assumed that at low temperature orbitals are frozen   
in a certain static pattern that optimizes both superexchange (SE) and 
orbital-lattice (JT) couplings.  
Such a classical approach has been used with a great success as a theoretical  
guide in studies of magnetism of transition-metal oxides. 
 
Recent experimental developments indicate, however, the limitations of  
this standard picture.  
It has been argued that quantum fluctuations of orbitals might sometimes be  
of crucial importance, hence quantum version of the orbital physics is needed. 
New concepts, such as three-dimensional orbital liquid in LaTiO$_3$
(Refs.~\onlinecite{KEI00} and onlinecite{KHA00})  
and one-dimensional orbital chains showing Heisenberg-like orbital dynamics  
in cubic vanadates (Ref.~\onlinecite{KHA01v}), have been proposed.  
It is not accidental that these ideas emerge from a study of titanates 
and vanadates  
having $t_{2g}^1$ and $t_{2g}^2$ electronic configurations, respectively.  
This is because of (i) large, threefold degeneracy, and  
(ii) a special rotational symmetry of $t_{2g}$ orbitals.  
Another crucial point is that (iii) the JT coupling 
is relatively weak for $t_{2g}$ systems. 
Indeed, JT-like elongation of octahedra on titanates and vanadates  
is much smaller compared with typical JT distortions 
in manganites with $e_g$ orbitals. 
One may therefore think that $t_{2g}$ orbital states are much less affected  
by electron-lattice coupling, and an intrinsic dynamics of coupled 
spin-orbital system governed by electronic superexchange 
interaction becomes the decisive factor in a first place.  
Effects of lattice distortions (which are always present) 
can then be accounted for in a next step.  
This point has actually been emphasized long ago by 
Kugel and Khomskii\cite{KUG82} indicating also very peculiar 
specific features of $t_{2g}$ spin-orbital models.\cite{KUG75}  
     
The aim of this paper is twofold. First, we study the
orbital state and orbital quantum dynamics in the ferromagnetic state
of superexchange model with $t_{2g}$ orbital degeneracy
in a cubic lattice. Second, we discuss the results in context of the
magnetic properties of
YTiO$_3$, a rare example 
of a ferromagnetic Mott insulator.\cite{KAT97,GOR82} 
Recent spin-wave data shows that the ferromagnetic (F) state of this material
is highly isotropic  
having the same exchange couplings in all cubic directions.\cite{ULR02}  
This is in sharp contrast with expectations from the conventional 
orbital ordering picture resulting commonly in a strong spatial 
anisotropy of the spin exchange bonds.\cite{GOO55,KAN59}  
This observation already indicates a rather unusual orbital 
state in YTiO$_3$. We would like also to understand a mechanism 
which stabilizes such a isotropic F-state in YTiO$_3$,  
having in mind that its sister compound, LaTiO$_3$ shows a completely 
different, antiferromagnetic (AF) state. Curiously enough, 
spin-exchange couplings  
in LaTiO$_3$ are also of cubic symmetry and spin gap is also small,  
\cite{KEI00} and these observations were understood in terms of 
fluctuating orbitals.\cite{KHA00,KHA01}  

We argue that AF and F states in $t_{2g}$ SE model 
are actually very close in energy and strongly compete. This is because 
in both states there are large-scale 
orbital fluctuations gaining almost the same amount of the  
superexchange energy. Yet the AF state is slightly lower  
because of an additional, composite spin-orbital fluctuation. 
However, an external parameter, namely, a larger distortion of Ti-O-Ti 
bonds due to a small size of Y-ion in case YTiO$_3$ induces an additional 
ferromagnetic coupling in all three directions, and stabilizes the spin F state.
This distortion induces also a gap for orbital excitations.
The orbital order pattern is very 
specific, and it supports exactly the same ferrocouplings in all three 
directions. We derive an effective spin Hamiltonian  
which includes effects of the relativistic spin-orbital coupling as well, 
and show that this Hamiltonian leads to  
spin-wave dispersion and spin gap consistent with experimental observations.   
Some of these results were presented in Ref.~\onlinecite{KHA02}.

The following part of this paper is structured as follows:  
Section II presents superexchange Hamiltonian in $t_{2g}^1$ Mott insulator;  
Sec.~III presents orbital ordering and fluctuations in the F state;  
Sec.~IV presents stabilization of the F state by Ti-O-Ti bond distortion;  
Sec.~V presents orbital gap induced by Ti-O-Ti bond distortion;  
Effective spin Hamiltonian and magnons are given in Sec.~(VI);  
Sec.~VII lists predictions for resonant x-ray scattering; 
Sec.~VIII gives orbital angular momentum contribution to the neutron scattering
cross-section; 
(IX)~Summary and discussion.  
The appendixes A-C contain some lengthy equations;  
Appendix D shows magnon softening by orbital fluctuations; 
Appendix E shows spin interactions in a previously reported orbital state for YTiO$_3$. 
    
\section{Hamiltonian} 
\label{sec:Hamiltonian} 
\subsection{Superexchange interaction in $t_{2g}$ orbital system}  
We start with a discussion of the model Hamiltonian.
In Mott insulators, the competition between kinetic and potential energies
is resolved in favor of strong correlations that lead to
a localized electron picture.
Charge localization is however not perfect: 
electrons still make virtual excursions to neighboring sites in order to
retain their kinetic energy at least partially. In terminology
of Mott-Hubbard insulators,\cite{IMA98} the zero-point charge motion 
is described as a high-energy virtual transition across the Mott gap.
Kinetic energy associated with these transitions
leads to superexchange interactions, which 
in orbitally degenerate systems strongly depends on the orbital structure.  
In general, it can be written as  
\begin{equation} 
H_{SE}^{ij} 
=  
\Bigl( {\vec S}_i \cdot {\vec S}_j +\frac{1}{4} \Bigl) 
\hat J_{ij}^{(\gamma)} ~ + ~\frac{1}{2}\hat K_{ij}^{(\gamma)} ~, 
\label{eq:original} 
\end{equation} 
where the orbital operators $\hat J_{ij}^{(\gamma)}$ and $\hat 
K_{ij}^{(\gamma)}$ depend on bond directions $\gamma(= a,b,c)$.  
In a $t_{2g}$ system like the titanates they are given by 
the following expressions:\cite{KHA01} 
\begin{eqnarray} 
\hat J_{ij}^{(\gamma)} & = &  J_{SE} \Bigl[ 
\case{1}{2}(r_1+r_2) A_{ij}^{(\gamma)}  
-\case{1}{3}(r_2-r_3) B_{ij}^{(\gamma)} \nonumber \\ 
& & \hspace{2.5em} -\case{1}{4}(r_1-r_2)(n_i+n_j)^{(\gamma)} \Bigr],  
\label{eq:Jgamma} \\ 
\hat K_{ij}^{(\gamma)} & = &  J_{SE} \Bigl[ 
\case{1}{2}(r_1-r_2) A_{ij}^{(\gamma)}  
+\case{1}{3}(r_2-r_3) B_{ij}^{(\gamma)} \nonumber \\ 
& & \hspace{2.5em} -\case{1}{4}(r_1+r_2)(n_i+n_j)^{(\gamma)} \Bigr] , 
\label{eq:Kgamma} 
\end{eqnarray} 
where $J_{SE} = 4t^2/U$.  
The coefficients  
$r_1=1/(1-3 \eta)$, $r_2=1/(1- \eta)$, and $r_3=1/(1+2 \eta)$ 
originate from the Hund's splitting of the excited $t_{2g}^2$ multiplet 
via $\eta=J_H/U$. Reference \onlinecite{MIZ96} gives  
$J_H \sim $ 0.64 eV and the multiplet averaged Coulomb interaction  
($=U- \frac{20}{9}J_H$) $\sim$ 4 eV, from which  
representative values $U \sim 5.4$~eV and $\eta \sim$~0.12 follow. 
 
The operators $A_{ij}^{(\gamma)}$, $B_{ij}^{(\gamma)}$,  
and $n_i^{(\gamma)}$ can conveniently be  
represented in terms of constrained particles (orbitons)  
$a_i$, $b_i$, $c_i$ with $n_{ia}+n_{ib}+n_{ic}=1$  
corresponding to $t_{2g}$ levels of $yz$, $xz$, $xy$ symmetry, respectively. 
(This notation is motivated by the fact that each $t_{2g}$ orbital is  
orthogonal to one of the cubic axes $a$,$b$,$c$.) 
Namely, 
\begin{eqnarray} 
A_{ij}^{(c)} & = & n_{ia}n_{ja}+n_{ib}n_{jb}  
+ a_i^\dagger b_i b_j^\dagger a_j 
+ b_i^\dagger a_i a_j^\dagger b_j, 
\label{eq:A_ab} 
\\ 
B_{ij}^{(c)} & = & n_{ia}n_{ja}+n_{ib}n_{jb} 
+ a_i^\dagger b_i a_j^\dagger b_j 
+ b_i^\dagger a_i b_j^\dagger a_j, 
\label{eq:B_ab} 
\\ 
n_i^{(c)} & = & n_{ia} + n_{ib}  
\label{eq:n_ab}  
\end{eqnarray} 
for the pair along the $c$ axis. 
Similar expressions are obtained for the exchange bonds along the axes 
$a$ and $b$, by replacing orbitons $(a, b)$ in Eqs.~(\ref{eq:A_ab})-(\ref{eq:n_ab})  
by $(b, c)$ and $(c, a)$ pairs, respectively.\cite{PAIR}  
Another useful representation of orbital exchange operators is via 
the angular momentum operators of $t_{2g}$ level,\cite{KUG75}  
using the  following relations: 
\begin{equation} 
l_x\!= i \bigl( c^\dagger b- b^\dagger c \bigr),  
l_y\!= i \bigl( a^\dagger c- c^\dagger a \bigr),  
l_z\!= i \bigl( b^\dagger a- a^\dagger b \bigr).  
\label{eq:angular} 
\end{equation} 
In terms of these angular momentum operators,  
$A_{ij}^{(\gamma)}$, $B_{ij}^{(\gamma)}$,   
and $n_i^{(\gamma)}$ are represented as 
\begin{eqnarray} 
A_{ij}^{(c)} & = & [(1-l_x^2)_i (1-l_x^2)_j + (l_x l_y)_i (l_y l_x)_j]  
+ [x \leftrightarrow y] , 
\label{eq:A_l} 
\\ 
B_{ij}^{(c)}  
& = & [(1-l_x^2)_i (1-l_x^2)_j + (l_x l_y)_i (l_x l_y)_j ] 
+ [x \leftrightarrow y] , 
\label{eq:B_l} 
\\ 
n_i^{(c)} & = & l_{i z}^2 . 
\label{eq:n_l} 
\end{eqnarray} 
Expressions of these operators for $a$ and $b$ bonds are given by replacing  
two component of the angular momentum $(l_x, l_y)$  
in Eqs.~(\ref{eq:A_l})-(\ref{eq:n_l}) with $(l_y, l_z)$ and $(l_z, l_x)$, 
respectively.  
Angular and quadrupole momentum representation of the $t_{2g}$ superexchange 
has recently been used also in Ref.~\onlinecite{ISH02}.   
In addition to Eqs.~(\ref{eq:A_ab})-(\ref{eq:B_ab}) 
and (\ref{eq:A_l})-(\ref{eq:B_l}),  
it is also useful to represent $A_{ij}^{(\gamma)}$ and $B_{ij}^{(\gamma)}$   
in terms of auxiliary orbital pseudospins:  
\begin{eqnarray} 
A_{ij}^{(\gamma)}  
&=& 2 \Bigl( \vec \tau_i \cdot \vec \tau_j + \frac{n_i n_j}{4} \Bigr)^{(\gamma)},  
\label{eq:A_tau} 
\\ 
B_{ij}^{(\gamma)}  
&=& 2 \Bigl( \vec \tau_i \otimes \vec \tau_j 
+ \frac{n_i n_j}{4} \Bigr)^{(\gamma)}.  
\label{eq:B_tau} 
\end{eqnarray} 
Here ${\vec \tau}_i^{(\gamma)}$ is  
a pseudospin one-half operating on the subspace of orbital doublet  
$(\alpha,\beta)^{(\gamma)}$  
active on a given $\gamma$-bond. Namely, pseudospin ${\vec \tau}_i^{(c)}$ 
operates on the subspace spanned by $(a,b)$ pair of orbitons, while  
${\vec \tau}_i^{(a)}$ and ${\vec \tau}_i^{(b)}$ act on 
$(b,c)$ and $(c,a)$ doublets, respectively. A symbol $\otimes$ denotes 
a product $\vec \tau_i \otimes \vec \tau_j =
\tau_i^z\tau_j^z + (\tau_i^+\tau_j^+ + \tau_i^-\tau_j^-)/2$.

\subsection{Ferromagnetic state} 

The ferromagnetic state of a Mott insulator is usually thought due to 
a particular orbital ordering that optimizes the intraatomic 
Hund's exchange of electrons in doubly occupied virtual states.
This is not the whole story, however. Neglect for a moment 
the Hund's coupling terms in Eqs.(\ref{eq:Jgamma}) and (\ref{eq:Kgamma})
(consider $\eta \rightarrow 0$ limit). 
The Hamiltonian obtains then the following structure: 
\begin{equation}
H_0=
J_{SE} \sum_{\left\langle ij \right\rangle}
2 \Bigl({\vec S}_i \cdot {\vec S}_j +\frac{1}{4} \Bigr) 
\Bigl( \vec \tau_i \cdot \vec \tau_j + 
\frac{1}{4}n_i n_j \Bigr)^{(\gamma)}~.
\label{Heta0}
\end{equation}
(The unessential energy shift, $-J_{SE}$, is not shown here). 
Regarding a single bond, one notices that spin coupling may be of either sign,
depending on the intersite orbital correlations. Singlet
correlations of orbital pseudospins tend to align spins ferromagnetically,
hence cooperating with Hund's rule effects. 
In systems with large, classical spins (e.g., vanadates), such a quantum 
orbital singlet controls the ground state.\cite{KHA01v,SHE02} 
In quantum spin one-half case of titanates, however, 
(spin triplet)$\times$(orbital singlet) and (spin singlet)$\times$(orbital triplet)
configurations are degenerate and compete. In a lattice, quantum
resonances between these configurations are possible.\cite{KHA00}
In general,
$t_{2g}$ superexchange Hamiltonian Eq.~(\ref{eq:original}) represents a highly 
frustrated many-body problem. We will return to the interplay
between antiferromagnetic and ferromagnetic states later on, while focusing now  
on the ferromagnetic state realized in YTiO$_3$.
 
In the spin saturated state, Eqs.~(\ref{eq:original})-(\ref{eq:Kgamma})
are simplified to:
\begin{equation} 
H_{orb} = -r_1 J_{SE} + \frac{1}{2} r_1 J_{SE} 
\sum_{\langle ij \rangle} A_{ij}^{(\gamma)} , 
\label{eq:Horb} 
\end{equation}
where $A_{ij}^{(\gamma)}$ is given by either of 
Eqs.(\ref{eq:A_ab}),(\ref{eq:A_l}), and (\ref{eq:A_tau}).
We consider the orbital order and dynamics in this Hamiltonian.
The effects of the dynamical coupling between spin excitations and orbitals
that is present in Eq.~(\ref{eq:original}) will also be discussed  
in the context of magnon spectra.
 
\section{Orbital ordering and excitations} 
\label{sec:order_excite} 
 
\subsection{Discussion of possible orderings} 
\label{subsec:possible} 
Even though spin as well as composite spin/orbital dynamics 
is ``switched off''
in Hamiltonian Eq.~(\ref{eq:Horb}), it still contains nontrivial physics.
 
It is useful to look at the structure of $H_{orb}$ from different
points of view. 
(i) On a given bond, the operator $A_{ij}^{(\gamma)}$
acts within a particular doublet of equivalent orbitals. 
Spin-like physics, that is the formation of orbital singlets
is therefore possible. 
(ii) On the other hand, interactions on different bonds
are competing: they involve different doublets, thus frustrating each other.
This brings about a Potts-model-like frustration, from which 
the high degeneracy of classical orbital configurations follows.  
(iii) Finally, we observe in $A_{ij}^{(\gamma)}$ a pseudospin 
$l=1$ interaction of pure biquadratic form 
[see Eq.~(\ref{eq:A_l})]. 
Would $\vec l$ be a classical vector, it could
change its sign at any site independently. 
Such a local (so-called $Z_2$) symmetry and
the associated degeneracy of the classical states tell us that
angular momentum ordering, if any, must be of pure quantum origin. 

The above points (i)-(iii) govern the underlying 
physics of the orbital Hamiltonian. We need to find such classical
states that provide best zero point energy when we switch on the
quantum fluctuations. In other words, certain classical orbital patterns
will be selected and stabilized by quantum effects via the order-from-disorder
mechanism.\cite{TSV95} Normally, these 
orderings are expected to be along symmetric orientations of the crystal
depending on symmetry of the underlying interactions.  
 
\begin{figure} 
\epsfxsize=0.8\columnwidth
\centerline{\epsffile{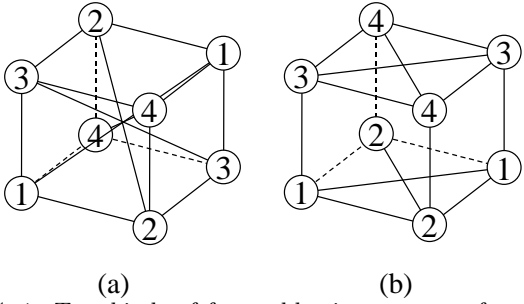}} 
\caption{Two kinds of four-sublattice structure for orbital orderings
in spin ferromagnetic $t_{2g}$ superexchange model.}  
\label{fig:number} 
\end{figure}

\begin{figure} 
\epsfxsize=0.9\columnwidth
\centerline{\epsffile{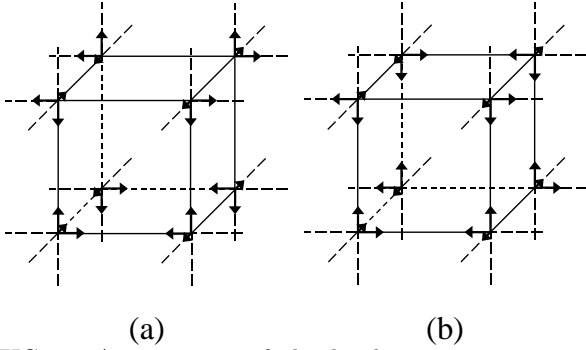}} 
\caption{Arrangement of the local quantization axes  
in states (a) and (b). Arrows indicate the quantization axes at each site,
and represent also a snapshot of local correlations of angular momentum:
on every bond, two out of three components of $\vec l$ are correlated
antiparallel.}  
\label{fig:2pattern} 
\end{figure}

By inspection of the global structure of 
$A_{ij}^{(\gamma)}$ [Eq.~(\ref{eq:A_l})],  
one observes that the non-cross terms, such as $(1-l_x^2)_i (1-l_x^2)_j$, 
are definitely positive.  
However, the cross terms, $(l_x l_y)_i (l_y l_x)_j$ and 
$(l_y l_x)_i (l_x l_y)_j$  
[which change the ``color'' of orbitals, see Eq.~(\ref{eq:A_ab})],  
can be made negative on {\it all the bonds simultaneously}, if  
(i) on every bond, two particular components of $\vec l_i$ and $\vec l_j$ 
are antiparallel, and  
(ii) remaining third components are parallel.  
For $c$ bonds the rule reads as: $l_{iz}l_{jz}$ and $l_{ix}l_{jx}$ are  
both negative, while $l_y$ components are parallel.  
(In terms of orbitons: $c_i$ and $c_j$ are in antiphase, $a_i$ and $a_j$ 
as well; but $b_i$ and $b_j$ have the same phase.)  
We find only two topologically different arrangements [called (a) and (b)],  
which can accommodate this curious mixture of ``2/3 antiferro'' plus 
``1/3 ferro'' correlations (see Figs.~\ref{fig:number} and \ref{fig:2pattern}).  
In the state (a), sublattice unit vectors are along 
the cubic diagonals [111],   
while in state (b), sublattice unit vectors are [110] and [002].  

For technical reasons, 
it is useful to introduce new quantization axes.  
This is done in two steps.  
First, we introduce local, sublattice specified 
quantization axes (see Fig.~\ref{fig:2pattern}): 
\begin{eqnarray} 
1 &:& (x,y,z) \rightarrow (x,y,z) , \nonumber \\ 
2 &:& (x,y,z) \rightarrow (-x,-y,z) , \nonumber \\ 
3 &:& (x,y,z) \rightarrow (-x,y,-z) , \nonumber \\ 
4 &:& (x,y,z) \rightarrow (x,-y,-z) .   
\label{xyz}
\end{eqnarray} 
After corresponding sign transformations of $l_{i\alpha}$ and orbitons,  
one obtains 
\begin{eqnarray} 
A_{ij}^{(c)} &=& n_{ia}n_{ja}+n_{ib}n_{jb}  
- a_i^\dagger b_i b_j^\dagger a_j - b_i^\dagger a_i a_j^\dagger b_j   
\label{eq:A_abnew} \\
&=& [(1-l_x^2)_i (1-l_x^2)_j - (l_x l_y)_i (l_y l_x)_j]  
+ [x \leftrightarrow y].   
\label{eq:A_lnew} 
\end{eqnarray} 
 
\begin{figure} 
\epsfxsize=0.8\columnwidth 
\centerline{\epsffile{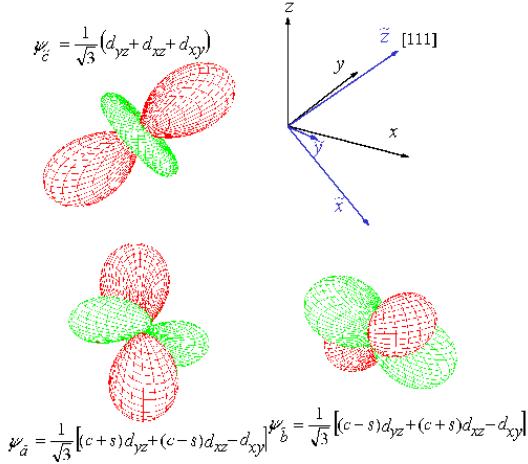}} 
\caption{(Color online). Orbitals in a new basis $\tilde x \tilde y \tilde z$
specified by transformation Eq.~(\ref{R}).}  
\label{fig:orbitals} 
\end{figure}

>From now on, a sublattice structure will not enter in the excitation 
spectrum. From the above observations it is also clear that all the 
components of $\vec l$ are equally needed  
to optimize all the three directions.  
We anticipate therefore that the cubic diagonals 
are ``easy'' (or ``hard'') axes for $\vec l$ fluctuations/orderings 
(recall that the Hamiltonian has no rotational symmetry for $\vec l$ vector).  
Therefore, it is convenient to further rotate the quantization axis  
so that new $z$-axis (denoted as $\tilde z$) corresponds to $[111]$ 
direction. This is done as follows:  
\begin{equation} 
\vec l_i = \hat R ~\vec{\tilde l}_i ,
\end{equation} 
where $\vec{\tilde l}_i =(\tilde l_{ix}, \tilde l_{iy}, \tilde l_{iz})$  
and $\hat R$ is given by  
\begin{eqnarray} 
\hat R = \frac{1}{\sqrt 3} 
\left( 
\begin{array}{ccc} 
c+s & c-s & 1 \\ 
c-s & c+s & 1 \\ 
-1 & -1 & 1 
\end{array} 
\right),  
\label{R}
\end{eqnarray} 
with $c=1/2$ and $s=\sqrt{3}/2$.  
Here, new $\tilde x$- and $\tilde y$-axes 
are taken to be symmetric with respect to the $[110]$ direction.  
Annihilation operators for constrained particles 
obey the same transformation,  
\begin{eqnarray} 
\left( \begin{array}{c} 
a \\ b \\ c  
\end{array} \right) 
= \hat R 
\left( \begin{array}{c} 
\tilde a \\ \tilde b \\ \tilde c  
\end{array} \right) . 
\label{eq:rotation} 
\end{eqnarray} 
Explicit expressions for the wave functions $\psi_{\tilde \alpha}$  
are obtained by reversing Eq.~(\ref{eq:rotation}) as follows:  
\begin{eqnarray} 
\psi_{\tilde a} &=& \frac{1}{\sqrt{3}}[(c+s)d_{yz}+(c-s)d_{xz}-d_{xy}], \nonumber \\ 
\psi_{\tilde b} &=& \frac{1}{\sqrt{3}}[(c-s)d_{yz}+(c+s)d_{xz}-d_{xy}], \nonumber \\ 
\psi_{\tilde c} &=& \frac{1}{\sqrt{3}}(d_{yz}+d_{xz}+d_{xy}).  
\label{eq:wave} 
\end{eqnarray} 
In Fig.~\ref{fig:orbitals}, we show schematic pictures of these orbitals.  
By construction, $\psi_{\tilde c}$ is symmetric with 
respect to the rotation around $[111]$ direction, having simply
$3\tilde z^2-r^2$ symmetry, while  
$\psi_{\tilde a}$ and $\psi_{\tilde b}$ are symmetric with respect to
$[110]$ axis. At the end, the orbital Hamiltonian in a rotated basis
obtains the following form 
(symbol ``tilde'' denoting rotated axes is implied for angular and
quadrupole operators below,
and constant energy shift is dropped out):  
\begin{equation}  
H_{orb}=\frac{1}{2}r_1 J_{SE}\sum_{\langle ij \rangle} A_{ij}^{(\gamma)}, 
\label{eq:Hrot2} 
\end{equation} 
with 
\begin{eqnarray} 
3A_{ij}^{(\gamma)} &=&  
\frac{2}{3} (1 -  Q_{iz} Q_{jz} )  
- \frac{1}{2} l_{iz} l_{jz} \nonumber \\ 
&& +\frac{1}{2}(Q_x T_{-1} + T_{-1} Q_x - T_0 T_1 -T_1 T_0)_{ij} \nonumber \\ 
&& +\frac{1}{2}(Q_x Q_x - T_0 T_0 + T_{-1} T_{-1})_{ij}^{(\gamma)} \nonumber \\ 
&& + \frac{2}{3} Q_{iz} (T_0 +cT_1)_j^{(\gamma)}  
+ \frac{2}{3} (T_0 +cT_1)_i^{(\gamma)} Q_{jz} \nonumber \\ 
&& +\frac{1}{2} l_{iz} (l_x + l_y)_j^{(\gamma)}  
+ \frac{1}{2}(l_x + l_y)_i^{(\gamma)} l_{jz}  
\nonumber \\ 
&& -\frac{1}{2}(l_x + l_y)_i^{(\gamma)} (l_x + l_y)_j^{(\gamma)},  
\label{eq:Hrot3} 
\end{eqnarray} 
where  
$Q_z$ and $Q_x$ represent the quadrupole moment operators with $e_g$ symmetry,  
$3z^2 - r^2$ and $x^2 -y^2$, respectively.  
$T_0=T_z$  
and $T_{\pm1}=T_y \pm T_x$,  
where    
$T_z$, $T_y$, and $T_x$ represent the quadrupole moment operators with $t_{2g}$ symmetry of 
$xy$, $xz$, and $yz$, respectively.  
These operators are expressed in terms of angular momentum and  
orbiton operators as follows:  
\begin{eqnarray} 
Q_z &=& \frac{1}{2} \bigl( \vec l^2 - 3 l_z^2 \bigr) 
=  n_{\tilde c} - \frac{1}{2} (n_{\tilde a} + n_{\tilde b}) , \nonumber \\ 
Q_x &=& l_x^2 - l_y^2  
= n_{\tilde b} - n_{\tilde a}  , \nonumber \\ 
T_z &=& l_x l_y + l_y l_x  
= -\bigl(\tilde a^\dag\tilde b +\tilde b^\dag\tilde a \bigr) , \nonumber \\ 
T_x &=& l_y l_z + l_z l_y  
= -\bigl(\tilde b^\dag\tilde c +\tilde c^\dag\tilde b \bigr) , \nonumber \\ 
T_y &=& l_x l_z + l_z l_x  
= -\bigl(\tilde c^\dag\tilde a +\tilde a^\dag\tilde c \bigr) . 
\label{eq:quadru} 
\end{eqnarray} 
$\gamma$-dependence of quadrupole moment operators is obtained by  
changing $l_{x,y}$ in Eq.~(\ref{eq:quadru}) to
\begin{eqnarray} 
l_x^{(\gamma)}= \left\{ \begin{array}{rl} 
-c l_x \pm s l_y, & \quad \mbox{for $\gamma = a (b)$} \\ 
l_x, & \quad \mbox{for $\gamma = c$}  
\end{array} \right. \nonumber \\ 
l_y^{(\gamma)}= \left\{ \begin{array}{rl} 
-c l_y \mp s l_x, & \quad \mbox{for $\gamma = a (b)$} \\ 
l_y. & \quad \mbox{for $\gamma = c$}  
\end{array} \right.  
\end{eqnarray} 
Explicit expressions for $Q_\alpha^{(\gamma)}$ and $T_\alpha^{(\gamma)}$  
are given in Appendix~\ref{sec:QTgamma}. 
It should be noted that,  
among eight operators, namely, five quadrupole moment and  
three angular momentum operators,  
only four operators are independent of each other  
because of the local constraint among orbiton operators.
 
Although it looks a bit complicated, the rotated Hamiltonian 
obtains a well-structured form. The first and second terms 
of Eq.~(\ref{eq:Hrot3}) represent  
``Ising''-like interaction for quadrupole moments and angular momenta.  
This part of the Hamiltonian stabilizes the ordering (condensation) 
of an appropriate orbiton.  
On the other hand, the other terms represent fluctuations 
of $Q_x$, $T_x$, $T_y$, $T_z$ and
transverse components of angular momenta $l_x$ and $l_y$.
These terms generate dispersion of the orbital excitations.  
 
Ordered states, promoted by the ``Ising'' part of interactions,
can be characterized by the quadrupole moment $Q=\langle Q_z \rangle$  
($Q$-order may couple to a lattice distortion of $D_{3d}$ symmetry),  
and the angular magnetic moment $m_l = \big\langle \tilde l_z \big\rangle$.  
We notice that the magnetic, $l_{iz}l_{jz}$ term in the first line of 
Eq.~(\ref{eq:Hrot3}) is generated by  
quantum commutation rules when we rotate $H_{orb}$;  
this makes explicit that the $Z_2$ symmetry is only a classical one  
and emphasizes the quantum origin of orbital magnetism.  

\begin{figure} 
\epsfxsize=0.95\columnwidth
\centerline{\epsffile{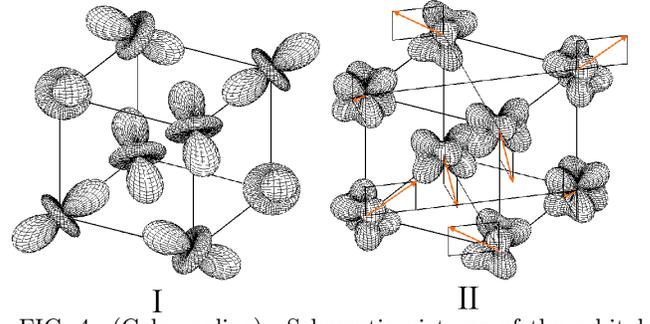}} 
\caption{(Color online). Schematic pictures of the orbital orderings.  
Left: Real orbital ordering I(a).
Right: Complex orbital ordering II(a).  
Here, the absolute values of the wave functions are presented. 
Arrows represent the directions of angular magnetic
momenta in the orbital magnetic state II(a).}  
\label{fig:order} 
\end{figure}

As it follows from the definition of $Q_z$, quadrupole ordering with finite
$Q$ but zero $l_z$ corresponds to a condensation of the  
$\tilde c$ orbiton. We call this solution state I. Classically,
$Q=1$ in this state.
On the other hand, condensation of the complex orbital  
$(\tilde a - i\tilde b)/ \sqrt 2$ generates a finite magnetic moment
($m_l=1$ classically), the state called II. The orbital
patterns in a classical states I and II are shown in Fig.~\ref{fig:order}.
We now focus on fluctuations of orbitals, 
and show that the excitation spectra are in fact identical in these states.
Moreover, we will obtain that the states I and II can smoothly be connected
by a continuous phase rotation of the condensate wave function.  
Noticing that an arbitrary cubic diagonal could be taken as $\tilde z$  
and having in mind also two structures in Fig.~\ref{fig:number},  
one obtains a multitude of degenerate states.  
This makes, in fact, all the orderings very fragile.  

\subsection{Orbital quadrupole order} 
\label{subsec:orb_quadru} 
\begin{figure} 
\epsfxsize=0.8\columnwidth 
\centerline{\epsffile{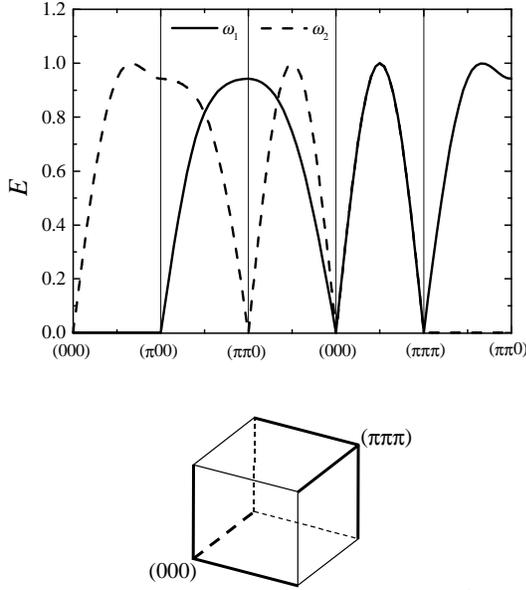}} 
\caption{Upper panel: Orbiton dispersions (in units of $r_1 J_{SE}$), 
obtained in a linear spin-wave approximation.  
Lower panel: Positions of soft modes are shown by the thick lines.}  
\label{fig:dispersion} 
\end{figure}

This state is driven by a condensation of $\tilde c$ orbital, that is,
ordering of the orbital $\psi_{\tilde c}$ in Eq.(\ref{eq:wave}).
To obtain a linear orbital wave Hamiltonian, we resolve a constraint
as $\tilde c=\tilde c^\dag=\sqrt{1- n_{\tilde a} - n_{\tilde b}}$,
and expand Eq.~(\ref{eq:Hrot3}) up to second order 
in $\tilde a$ and $\tilde b$. The result is (in units of $r_1 J_{SE}$):
\begin{eqnarray} 
H_{OW} &=& \sum_i (n_{i\tilde a} + n_{i\tilde b}) \nonumber \\ 
&+&\frac{1}{2z}\sum_{\langle ij \rangle}  
\Bigl[T_{i \; -1}^{(\gamma)} T_{j \; -1}^{(\gamma)}  
- (l_x + l_y)_i^{(\gamma)} (l_x + l_y)_j^{(\gamma)} \Bigr] ,  
\label{eq:How1} 
\end{eqnarray} 
where $z=6$, and 
\begin{eqnarray} 
&T_{\pm1} = -\bigl(\tilde a^\dag +\tilde a \bigr) \mp \bigl(\tilde b^\dag +\tilde b \bigr) ,& \nonumber \\ 
&l_x = i \bigl(\tilde b -\tilde b^\dag \bigr) , \quad 
l_y = i \bigl(\tilde a^\dag -\tilde a \bigr) .& 
\label{eq:Tll} 
\end{eqnarray} 
In a momentum space, this linearized Hamiltonian reads as  
\begin{eqnarray} 
H_{OW} &=& \sum_{\vec k} \Bigl[ 
n_{\tilde a \vec k} + n_{\tilde b \vec k}  
+\frac{1}{2} (\gamma_1 + \gamma_2)  
\bigl(\tilde a_{\vec k}^\dag\tilde a_{-\vec k}^\dag +\tilde a_{\vec k}\tilde a_{-\vec k} \bigr) \nonumber \\ 
&& \hspace{2em}  
+\frac{1}{2} (\gamma_1 - \gamma_2)  
\bigl(\tilde b_{\vec k}^\dag\tilde b_{-\vec k}^\dag +\tilde b_{\vec k}\tilde b_{-\vec k} \bigr) 
\nonumber \\ 
&& \hspace{2em} 
- \gamma_3 \bigl(\tilde a_{\vec k}^\dag\tilde b_{-\vec k}^\dag +\tilde a_{\vec k}\tilde b_{-\vec k} \bigr) 
 \Bigr] ,  
\label{eq:How2} 
\end{eqnarray} 
where $\gamma_1$, $\gamma_2$, and $\gamma_3$ are defined as  
$\gamma_1 = (c_x + c_y + c_z)/3$, $\gamma_2 = \sqrt{3}(c_y - c_x)/6$, and  
$\gamma_3 = (2c_z -c_x -c_y)/6$, respectively, 
with $c_{\alpha}=\cos k_{\alpha}$.  
The Hamiltonian is diagonalized by using Bogoliubov transformation 
(see for details Appendix~\ref{sec:Bogoliubov}). One obtains:  
\begin{eqnarray} 
H_{OW} = \sum_{\vec k} 
\Bigl( \omega_{1 \vec k} \alpha_{1 \vec k}^\dag \alpha_{1 \vec k} 
+\omega_{2 \vec k} \alpha_{2 \vec k}^\dag \alpha_{2 \vec k} \Bigr) +E_0 ,  
\label{eq:diagonal} 
\end{eqnarray} 
where $\omega_{1 \vec k}=\sqrt{1-(\gamma_1 + \kappa)^2}$ and  
$\omega_{2 \vec k}=\sqrt{1-(\gamma_1 - \kappa)^2}$ with  
$\kappa = \sqrt{\gamma_2^2 + \gamma_3^2}$.  
The dispersion relations of orbitons are presented 
in Fig.~\ref{fig:dispersion}.  
Orbital excitations are characterized by the flat dispersion with zero energy  
along $(0,0,k_z)$, $(\pi,\pi,k_z)$ and their equivalent directions. 
As discussed later, interaction effects open the gap along $(\pi,\pi,k_z)$
and equivalent ones, while zero modes along 
$(k_x,0,0)$, $(0,k_y,0)$, and $(0,0,k_z)$ are protected by the underlying
symmetry of the model.
 
A constant $E_0$ in Eq.~(\ref{eq:diagonal}) represents 
the energy gain due to the quantum fluctuations. It is given by 
\begin{eqnarray} 
E_0 = {1 \over 2}\sum_{\vec k}(\omega_{1 \vec k} + \omega_{2 \vec k})-1  
= -0.214~~ (r_1 J_{SE}).  
\label{eq:energy} 
\end{eqnarray}  

We may compare this result with ground-state energy of 
the orbital disordered AF state:  
$E_0=-0.33 \frac{r_1 + r_2}{2} J_{SE}$,\cite{KHA00}  
where the result of Ref.~\onlinecite{KHA00} is
corrected for the finite values of $\eta$. 
For realistic values of the Hund's coupling, say $\eta = J_H/U = 0.12$,
this gives $E_0=-0.285$ (in units of $r_1 J_{SE}$) in AF state. 
It is noticed that ferromagnetic and AF states are almost degenerate.
Still, the ferromagnetic state is higher than the AF state, so its
stabilization in YTiO$_3$ requires an additional effects as discussed
in Sec.~\ref{sec:why}.  
 
\begin{figure} 
\epsfxsize=0.9\columnwidth 
\centerline{\epsffile{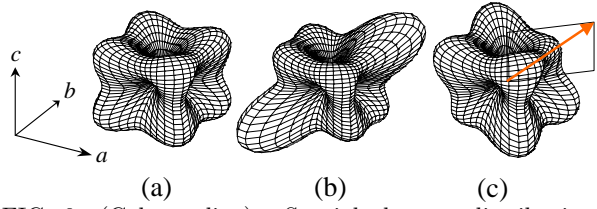}} 
\caption{(Color online). Spatial electron distribution in different states. 
(a)Disordered state, $n_a = n_b = n_c = \frac{1}{3}$.  
(b)Quadrupole ordered state, $n_{\tilde c} = 0.46$ and  
$n_{\tilde a} = n_{\tilde b} = 0.27$. 
(c)Orbital magnetic state, $n_{\bar a} = 0.46$ and  
$n_{\bar b} = n_{\tilde c} = 0.27$. 
Arrow shows the direction of angular momentum.}  
\label{fig:density} 
\end{figure}

Due to the flat mode, one may expect strong orbital fluctuations
in the ground state. Indeed,  
number of the excited bosons $\tilde a$ and $\tilde b$ is large 
even at $T=0$: 
\begin{eqnarray} 
\langle n_{i\tilde a} + n_{i\tilde b} \rangle = -1  
+ \frac{1}{2} \sum_{\vec k} \biggl( {1 \over \omega_{1 \vec k}} + {1 \over \omega_{2 \vec k}} \biggr) = 0.54.  
\label{eq:nab} 
\end{eqnarray} 
This reduces the condensate density 
to $\langle n_{i\tilde c} \rangle =0.46$. 
Consequently, the quadrupole order parameter is obtained to be rather
small: $Q=0.19$. 
Reduction of quadrupole order $Q$ implies that electron density
is much less anisotropic than that shown for
the classical state in Fig.~\ref{fig:order}(a).
Including fluctuation effects, that is, finite population
of $\tilde a$ and $\tilde b$ orbitals, 
electron density at site 1 is given by  
$\rho_1 (\vec r) = 
n_{\tilde c} \psi_{\tilde c}^2 + n_{\tilde a} \psi_{\tilde a}^2  
+ n_{\tilde b} \psi_{\tilde b}^2$.  
Using Eq.~(\ref{eq:wave}), one then finds  
\begin{eqnarray} 
\rho_1 (\vec r) &=&  
\frac{1}{3} \bigl( d_{yz}^2 + d_{xz}^2 + d_{xy}^2 \bigr) \nonumber \\ 
&& + \frac{2}{3} Q 
(d_{yz} d_{xz} + d_{yz} d_{xy} + d_{xz} d_{xy}) .  
\label{eq:density} 
\end{eqnarray}  
Electron density at other sites is given by a similar 
equation,  
where the second, of $t_{2g}$ symmetry term, is different 
for different sublattices.  
Namely, it is $(d_{yz} d_{xz} - d_{yz} d_{xy} - d_{xz} d_{xy})$ for site 2,   
$(- d_{yz} d_{xz} + d_{yz} d_{xy} - d_{xz} d_{xy})$ for site 3,  
and $(- d_{yz} d_{xz} - d_{yz} d_{xy} + d_{xz 
} d_{xy})$ for site 4.  
In Fig.~\ref{fig:density}(b), we present the 
electron distribution given by Eq.~(\ref{eq:density}).  
For comparison, we show in Fig.~\ref{fig:density}(a) the electron distribution  
where $\tilde a$, $\tilde b$, and $\tilde c$ are equally occupied.  
At finite $Q$, the electron density $\rho_1 (\vec r)$ is slightly 
elongated along $[111]$ direction.  
Thus, we expect the quadrupole ordered state to be further stabilized 
by the electron lattice coupling,  
although this coupling is expected to be weak for $t_{2g}$ orbitals.  

The anomalous reduction of the order parameter is due to the highly frustrated 
nature of the interactions in Eq.~(\ref{eq:Horb}). 
A special, non-spin-like feature of all orbital models is that 
orbitals are bond selective, resulting in a pathological 
degeneracy of classical states. This leads to soft modes [observe that
$\omega_{1,2 \vec k}$ is just flat along ($0,0,\pi$) and equivalent
directions]. These soft modes have their origin in special symmetry
properties of the $t_{2g}$ orbital model Eq.~(\ref{eq:Horb}), which result
in conservation laws with important consequences.
Namely, the total number of orbitals of each ``color'' ($a,b,c$) 
are conserved during superexchange process, as can easily be
seen from Eq.(\ref{eq:A_ab}). Moreover, as $t_{2g}$-orbitals can hop only
along two directions [say, $xy$-orbital motion is restricted to ($ab$)
planes], the orbital number is conserved on each plane separately.
Formally, these conservation rules are reflected by a possibility
of uniform phase transformation of orbiton operators, e.g.,
$a \rightarrow a ~exp(i\phi_a)$, {\it etc}., which leaves the orbital
Hamiltonian invariant. These continuous symmetries are spontaneously broken
in above ordered states. The breaking of continuous
symmetry is usually followed by the generation of gapless Goldstone modes.
This is precisely what happens in the $t_{2g}$ orbitally degenerate model.
In fact, soft modes obtained above have two-dimensional (2D) feature 
(stemming from 2D geometry of the $t_{2g}$ orbital hoppings). 
As a result, long range orbital order is possible only at zero temperature,
just like in 2D Heisenberg models. Formally, this is manifested
as a divergence (as $\ln \frac{1}{T}$) of the number of thermally 
excited orbitons, 
$\langle n_{\tilde a} + n_{\tilde b} \rangle$, if one attempts to 
calculate this quantity at finite temperature, 
including the Bose population factor in Eq.(\ref{eq:nab}).
Soft modes will be discussed in more detail later on.
 
\subsection{Orbital magnetic order} 
\label{subsec:orb_mag} 
In order to describe the magnetic ordering (denoted above by state II), 
let us introduce orbital states  
\begin{eqnarray} 
\bar a = \frac{1}{\sqrt{2}} (\tilde a - i \tilde b) , \quad 
\bar b = \frac{1}{\sqrt{2}} (\tilde a + i \tilde b) .  
\end{eqnarray} 
In these states, angular momentum has eigenvalues $\tilde l_z = \pm 1$
(on local axes), respectively. 
A condensation of $\bar a$ leads therefore to 
the magnetic ordering (with $\tilde l_z = 1$ classically), and it is
associated with ordering of the complex orbital:
\begin{equation}
\psi_l = \frac{1}{\sqrt{3}} 
\Bigl\{ d_{yz} e^{i \frac{\pi}{3}}
+ d_{xz} e^{-i \frac{\pi}{3}} - d_{xy} \Bigr\}.  
\label{psi_l}
\end{equation}
On global axes,
this order is noncollinear, as shown in Fig.~\ref{fig:order}(b).
We consider fluctuations of this state. 
Using the relation $\bar a = \bar a^\dag = 
\sqrt{1- n_{\bar b} - n_{\tilde c}}$ and expanding interactions 
in Eq.~(\ref{eq:Hrot3}) up to second order in $\bar b$ and $\tilde c$,
we obtain the following linearized Hamiltonian:  
\begin{eqnarray} 
H_{OW} &=& \sum_i (n_{i \bar b} + n_{i\tilde c}) \nonumber \\ 
&&+ \frac{1}{2z} \sum_{\langle ij \rangle}  
\Bigl[Q_{i x} T_{j \; -1} + T_{i \; -1} Q_{j x} - T_{i 0} T_{i 1} - T_{i 1} T_{j 0} \nonumber \\ 
&& \hspace{4em} 
+ Q_{i x}^{(\gamma)} Q_{j x}^{(\gamma)} - T_{i 0}^{(\gamma)} T_{j 0}^{(\gamma)}  
+ T_{i \; -1}^{(\gamma)} T_{j \; -1}^{(\gamma)} \nonumber \\ 
&& \hspace{4em} 
- (l_x + l_y)_i^{(\gamma)} (l_x + l_y)_j^{(\gamma)} \Bigr].   
\label{eq:HrotIm} 
\end{eqnarray} 
Here, the operators $l$, $Q$, and $T$ are linear functions of $\bar b$ and
$\tilde c$:  
\begin{eqnarray} 
&l_x = -\frac{1}{\sqrt{2}} \bigl(\tilde c^\dag +\tilde c \bigr), \quad 
l_y = -\frac{i}{\sqrt{2}} \bigl(\tilde c^\dag -\tilde c \bigr),& \nonumber \\ 
&Q_x= - \bigl( \bar b^\dag + \bar b \bigr), \quad 
T_0= -i(\bar b^\dag - \bar b),& \nonumber \\ 
&T_{\pm 1} = -\frac{1}{\sqrt{2}} \bigl[ (1 \mp i)\tilde c +(1 \pm i)\tilde c^\dag \bigr] .& 
\end{eqnarray} 
We introduce now new operators $\alpha,\beta$:
\begin{eqnarray} 
\alpha = \frac{1}{\sqrt{2}}(\bar b + e^{i \varphi}\tilde c),  
\quad 
\beta = \frac{1}{i\sqrt{2}}(\bar b - e^{i \varphi}\tilde c),  
\label{eq:alpha} 
\end{eqnarray} 
with $\varphi = \pi/4$. Remarkably, after this transformation
the linearized Hamiltonian obtains in a momentum space  
the same form as Eq.~(\ref{eq:How2}),  
where $\tilde a, \tilde b$ are just replaced by $\alpha,\beta$, 
and $\gamma_2$ and $\gamma_3$ are interchanged.  
Exchange of $\gamma_2$ and $\gamma_3$ does not affect  
the excitation spectrum, however, as they enter in $\omega_{1,2 \vec k}$ 
via the parameter $\kappa = \sqrt{\gamma_2^2 + \gamma_3^2}$ only.
Thus, the ground-state energy in the orbital magnetic state 
$E_0$ is given again by Eq.~(\ref{eq:energy}),  
so the states I and II are degenerate even on quantum level. 
Similarly, the number of out of
condensate bosons is also obtained from Eq.~(\ref{eq:nab}). This gives  
$\langle n_\alpha + n_\beta \rangle = 0.54$, and values for  
the angular and quadrupole momentum order parameters follow:  
\begin{eqnarray} 
m_l&=&1 -\frac{3}{2} \langle n_\alpha + n_\beta \rangle = 0.19,  
\nonumber \\
Q &=& -\frac{1}{2} + \frac{3}{4} \langle n_\alpha + n_\beta \rangle = -0.095.
\label{ml}
\end{eqnarray} 
Electron-density distribution in the state II, 
given by Eq.~(\ref{eq:density}) with above value of $Q$, 
is shown in Fig.~\ref{fig:density}~(c).  
Electron cloud in the magnetic state is almost of cubic symmetry,
being just slightly contracted along $[111]$ 
direction (opposite to that in the state I).  
Thus the energy gain from the orbital lattice coupling  
is smaller in this state.  
In principle, orbital magnetic order could be supported 
by a relativistic spin-orbital coupling; however, in the noncollinear 
state driven by the superexchange interaction, uniform component of 
the orbital moment is zero [see Fig.~\ref{fig:order}(b)], 
hence the coupling to the spin ferromagnetism vanishes in linear order. 
The spin-orbital energy gain in a second order is possible though, 
via the canting of spins towards orbital magnetic pattern.
    
\subsection{Soft modes: density-phase formulation} 
\label{subsec:phases} 
Having obtained an identical excitation spectrum for states with apparently
different ground-state condensates, we would like to unify these states.
It is convenient to use a different approach, that is, the 
density-phase formulation by Popov (Ref.~\onlinecite{POPOV}), 
nowadays called ``radial gauge''. 
This formalism is particularly useful also to clarify the physical
origin of the soft modes obtained above.
In the radial gauge,  
density and phase degrees of freedom of the constrained particles
are emphasized. We work in a basis obtained by first transformation, 
Eq.(\ref{xyz}), and represent the orbiton operators entering
in Eq.(\ref{eq:A_abnew}) as follows:   
\begin{eqnarray} 
\alpha_i=\sqrt{\rho_{i\alpha}} e^{i \theta_{i\alpha}} \quad (\alpha=a,b,c). 
\end{eqnarray} 
Further, the density and phase degrees of freedom are parametrized as  
\begin{eqnarray} 
\rho_{i \, a (b)} &=& \rho_0 + 
\frac{2}{3} ( c r_i \pm s \lambda_i), \nonumber \\ 
\rho_{ic} &=& \rho_0 - \frac{2}{3} r_i,  
\label{rho}
\end{eqnarray} 
and  
\begin{eqnarray} 
\theta_{i \, a (b)} &=& \Omega_i + c \varphi_i \pm s \theta_i , \nonumber \\ 
\theta_{ic} &=& \Omega_i -\varphi_i ,  
\label{phase}
\end{eqnarray} 
respectively.  
Here, $\rho_0 (=1/3)$ is an average electron density on each orbital.  
The phase $\Omega_i$, common to all the three orbitons,
can as usually be absorbed by the constraint field, while the local
constraint itself is explicitly resolved by 
parametrization (\ref{rho}).   
The physically active degrees of freedom are therefore 
$r$ and $\lambda$ fields for the amplitude fluctuations,  
and $\theta$ and $\varphi$ for the phase fluctuations.  
We recall that the coefficients $c=1/2, s=\sqrt{3}/2$. 

To start with, let us neglect for a moment the amplitude fluctuations, 
and focus on the phase dependence of the classical condensate wave function.
In terms of phases in Eq.(\ref{phase}), it is written as follows 
(up to unessential overall phase factor): 
\begin{eqnarray} 
\psi(\theta,\varphi)\!=\!\sqrt{\rho_0} 
\Bigl\{ d_{yz} e^{i (3c \varphi + s\theta)}
+ d_{xz} e^{i (3c \varphi - s\theta)}
+ d_{xy} \Bigr\}.  
\label{condensate}
\end{eqnarray} 
Here, we suppressed site dependence of the phases $\varphi,\theta$,
discarding for a while slow space variations of the condensate.
Now, it is noticed that the quadrupole and magnetic orderings
[see Eqs.(\ref{eq:wave}) and (\ref{psi_l})] do 
follow from Eq.(\ref{condensate}) when $\varphi=\theta=0$, 
and $\varphi=\frac{\pi}{3c}$, $\theta = \frac{\pi}{3s}$,
respectively. Next observation is the orbital ``color'' conservation
rule in the ferromagnetic state. In the radial gauge, it is evident
from Eq.(\ref{eq:A_abnew}), that the interactions do depend on the
{\it difference} of the orbiton phases only, 
that is on $\theta_{ia} - \theta_{ja}$, {\it etc}., 
so we can uniformly rotate the condensate function (\ref{condensate})   
by arbitrary phases $\varphi,\theta$ with no energy cost.
By such rotations, we can in fact mix quadrupole (state I)
and magnetic (state II) orderings. Slow phase rotation of
the condensate is precisely the origin of the soft modes obtained above.
Because of the two dimensionality of $t_{2g}$ orbitals, the phases
can spontaneously be fixed at zero temperature only. Of course, 
orbital-lattice and/or spin-orbital couplings may fix the phases, thus
selecting a particular state even at finite $T$. 

We now turn to the excitations of the model in density-phase formulation.
$A_{ij}^{(\gamma)}$ in Eq.~(\ref{eq:A_abnew}) is expressed as  
\begin{eqnarray} 
A_{ij}^{(c)}&=&\bigl(\sqrt{\rho_{ia}\rho_{ja}} -  
\sqrt{\rho_{ib}\rho_{jb}}\bigr)^2  \nonumber \\
&+&2 \sqrt{\rho_{ia}\rho_{ja}}\sqrt{\rho_{ib}\rho_{jb}}
\bigl\{1 - \cos \bigl(\phi_i^{(c)} - \phi_j^{(c)} \bigr) \bigr\}, 
\label{eq:Ac_theta} 
\end{eqnarray} 
where $\phi_i^{(c)} = \theta_{ia} - \theta_{ib}$.  
$A_{ij}^{(\gamma)}$ for $\gamma = a$ and $b$ bonds are given by replacing  
$(a,b)$ in Eq.~(\ref{eq:Ac_theta}) with $(b,c)$ and $(c,a)$, respectively.
In terms of the relevant phase degrees of freedom $\varphi$ and $\theta$,
we obtain  
\begin{eqnarray} 
\phi_i^{(\gamma)}= \left\{ \begin{array}{rl} 
\sqrt{3} (-c \theta_i \pm s \varphi_i ) , 
& \quad \mbox{for $\gamma = a (b)$} \\ 
\sqrt{3} \theta_i.  
& \quad \mbox{for $\gamma = c$}  
\end{array} \right.  
\end{eqnarray} 
The density operators $\rho_{i \alpha}$ are functionals of 
the $r_i$ and $\lambda_i$ fields. We may expand now the operator 
$A_{ij}^{(\gamma)}$ in terms of the amplitude and phase
variables $r, \lambda, \varphi, \theta$. Keeping quadratic only terms
in the expansion, one arrives at the following linearized
Lagrangian for the phase-amplitude fluctuations   
\begin{eqnarray} 
L_{orb}=\sum_{i,w} w(\theta_{i,w} \lambda_{i,-w}+\varphi_{i,w}r_{i,-w})
+ H_{\theta \varphi} + H_{\lambda r}, 
\label{eq:Lorb} 
\end{eqnarray} 
where $\omega$ is the Matsubara frequency. The first term in this equation
originates from the time derivative, kinematic part of the Lagrangian
$-\sum_{i,\alpha}\alpha_i^{\dagger}\frac{\partial}{\partial \tau}\alpha_i$,
and produces dynamical coupling between the density and phase variables.  
$H_{\theta \varphi}$ represents the phase fluctuations, and it is 
obtained from the expansion of the second term in Eq.~(\ref{eq:Ac_theta}):
\begin{eqnarray} 
H_{\theta \varphi} = \frac{1}{2} \rho_0^2 \sum_{\langle ij \rangle}  
\Bigl(\phi_i^{(\gamma)} - \phi_j^{(\gamma)} \Bigr)^2 . 
\label{eq:Htp} 
\end{eqnarray} 
The fluctuations of the condensate density about $\rho_0$
are controlled by the Hamiltonian $H_{\lambda r}$, which in a linear
approximation reads as   
\begin{eqnarray} 
H_{\lambda r} = \frac{9}{8}\rho_0^2 \sum_{\langle ij \rangle} 
\Bigl(\delta_i^{(\gamma)} + \delta_j^{(\gamma)} \Bigr)^2 .
\label{eq:Hrl} 
\end{eqnarray} 
Here, $\delta_i^{(\gamma)}$ is the difference between the electron 
densities on orbitals that are active on a given direction:  
$\delta_i^{(\gamma)} = \rho_{i\alpha}^{(\gamma)} - \rho_{i\beta}^{(\gamma)}$. 
Using parametrization (\ref{rho}), they are expressed via $r$ and $\lambda$
fields as follows: 
\begin{eqnarray} 
\delta_i^{(\gamma)}= \left\{ \begin{array}{rl} 
\frac{2}{\sqrt{3}} (-c  \lambda_i \pm s  r_i ),  
& \quad \mbox{for $\gamma = a (b)$} \\ 
\frac{2}{\sqrt{3}} \lambda_i. & \quad \mbox{for $\gamma = c$}  
\end{array} \right.  
\end{eqnarray} 
 
In a momentum space, $H_{\theta \varphi}$ and $H_{\lambda r}$ are represented as follows: 
\begin{eqnarray} 
H_{\theta \varphi} 
&=&\sum_{\vec q}\bigl\{\frac{1}{2}a_\theta |\theta_{\vec q}|^2  
+ \frac{1}{2} a_\varphi |\varphi_{\vec q}|^2 - \gamma_2 \varphi_{\vec q} \,\,
\theta_{-\vec q}\bigr\} , 
\label{ham_tv} \\ 
H_{\lambda r} &=& 
\sum_{\vec q}\bigl\{\frac{1}{2} a_\lambda |\lambda_{\vec q}|^2  
+ \frac{1}{2} a_r |r_{\vec q}|^2 +  
\gamma_2 \lambda_{\vec q} \,\, r_{-\vec q}\bigr\},  
\label{ham_lr}
\end{eqnarray} 
where $a_\theta = 1- (\gamma_1 + \gamma_3)$, 
$a_\varphi = 1 - (\gamma_1 - \gamma_3)$,  
$a_\lambda = 1+ (\gamma_1 + \gamma_3)$, and  
$a_r = 1 + (\gamma_1 - \gamma_3)$. 

Equations (\ref{eq:Lorb}) and (\ref{ham_tv}), (\ref{ham_lr}) determine
the orbital dynamics in a harmonic approximation, 
which is equivalent to the previous linear orbital wave approach. 
Indeed, the obtained quadratic form 
in Eq.(\ref{eq:Lorb}) can easily be diagonalized giving exactly 
the same excitation spectrum, that is 
$\sqrt{1-(\gamma_1 \pm \kappa)^2}$ ~found in the preceding section.
In addition, the origin of zero energy excitations can
clearly be identified now. In a classical limit [neglect dynamical
term in Eq.(\ref{eq:Lorb})], the quadratic forms 
$H_{\theta \varphi}$ (\ref{ham_tv}) and $H_{\lambda r}$ (\ref{ham_lr}) 
can be diagonalized separately resulting in normal modes with energies
$\omega_{\theta}^{\pm}(\vec k)=(1- \gamma_1 \pm \kappa)$ in phase sector, and 
$\omega_{\rho}^{\pm}(\vec k)= (1+ \gamma_1 \pm \kappa)$ 
for the amplitude variables.
$\omega_{\theta}^{\pm}$ vanishes on lines $(k_x,0,0),(0,k_y,0),(0,0,k_z)$
(see Fig.~\ref{fig:dispersion}). Therefore, zero-energy excitations on these
lines do correspond to uniform phase rotations of orbitons on different
planes as discussed before. On the other hand, the normal modes 
$\omega_{\rho}^{\pm}(\vec k)$ that are associated with the density 
of orbital occupancies possess zero lines at
$(k_x,\pi,\pi),(\pi,k_y,\pi),(\pi,\pi,k_z)$.
This reflects softness of the {\it staggered} fluctuations of orbitals
[notice also that a uniform component, that is, $\delta_u^{(\gamma)} = 
\delta_i^{(\gamma)}+\delta_j^{(\gamma)}$, only enters in Eq.(\ref{eq:Hrl})]. 
However, such soft modes in the amplitude sector are not protected by  
physical conservation rules; therefore, they are expected to acquire 
a finite mass due to interaction effects that go beyond linear orbital wave 
approximation. To see this, one should consider unharmonic terms in the
expansion of Eq.(\ref{eq:Ac_theta}). Most relevant term in that expansion
is the interaction between the staggered fluctuations, that is, 
$\delta_s^{(\gamma)} = \delta_i^{(\gamma)}-\delta_j^{(\gamma)}$, with
uniform components $\delta_u^{(\gamma)}$. When such a term is kept, 
Eq.(\ref{eq:Hrl}) is replaced by
\begin{eqnarray} 
H_{\lambda r} &=& \frac{9}{8}\rho_0^2 \sum_{\langle ij \rangle} 
\Bigl\{\delta_u^2 + 
\frac{1}{(4\rho_0)^2}\delta_s^2 \delta_u^2 \Bigr\}^{(\gamma)}
\nonumber \\
&\simeq& \frac{9}{8}\rho_0^2 \sum_{\langle ij \rangle} 
\Bigl\{ (1 +\varepsilon_s) \delta_u^2 + 
\varepsilon_u \delta_s^2 \Bigr\}^{(\gamma)}. 
\label{Hrlnew} 
\end{eqnarray} 
The Hartree decoupling is applied here to the interaction term, with  
$\varepsilon_{s,u}=(1/4\rho_0)^2 \langle \delta_{s,u}^2 \rangle$. These
expectation values are finite due to the presence of quantum fluctuations
in the ground state that are particularly enhanced in a staggered channel. 
In a momentum space, the above equation reads as
\begin{equation} 
H_{\lambda r}=Z_{\varepsilon}
\sum_{\vec q}\bigl\{\frac{1}{2} \tilde a_\lambda |\lambda_{\vec q}|^2  
+ \frac{1}{2} \tilde a_r |r_{\vec q}|^2 + 
\gamma_2^{\varepsilon} \lambda_{\vec q} \,\, r_{-\vec q}\bigr\},
\label{ham_Z}
\end{equation} 
where 
$\tilde a_\lambda = 1+ (\gamma_1^{\varepsilon} + \gamma_3^{\varepsilon})$ 
and  $\tilde a_r = 1 + (\gamma_1^{\varepsilon} - \gamma_3^{\varepsilon})$.  
$\gamma_n^\varepsilon$ $(n=1,2,3)$ are defined as  
$\gamma_n^\varepsilon = (1- 2\varepsilon)\gamma_n$ with $\varepsilon = 
\varepsilon_u/Z_{\varepsilon}$, and 
$Z_{\varepsilon}=(1+\varepsilon_u+\varepsilon_s)$ is 
an overall rescaling factor. 
 
\begin{figure} 
\epsfxsize=0.8\columnwidth 
\centerline{\epsffile{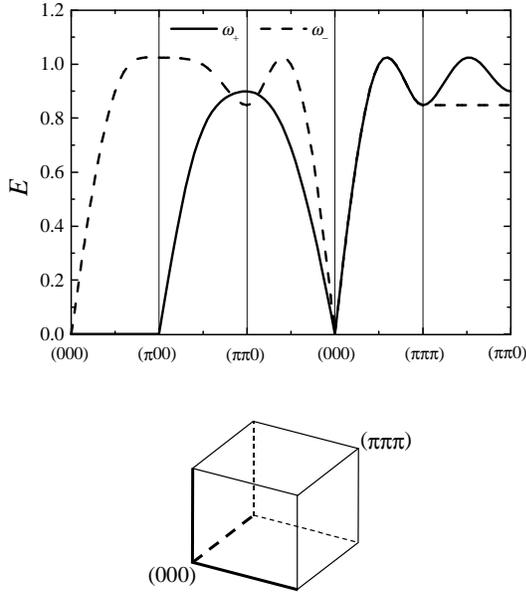}} 
\caption{Orbiton dispersions corrected by interaction effects
[Eq.~(\ref{omegapm})].   
Energy is given in units of 
$Z_{\varepsilon}^{1/2} r_1 J_{SE} \simeq 1.82~r_1 J_{SE}$.  
The amplitude fluctuations open the excitation gap around $(\pi \pi \pi)$,  
while there still remain gapless Goldstone modes at  
positions, indicated by the thick lines in the lower panel.}  
\label{fig:dis_gap} 
\end{figure}

Using now $H_{\lambda r}$ given by Eq.(\ref{ham_Z}) 
in $L_{orb}$ [Eq.~(\ref{eq:Lorb})],  
one obtains finally the following two eigenfrequencies:  
\begin{equation} 
\omega_\pm (\vec k) 
=\Bigl\{1-(1-2\varepsilon)(\gamma_1 \pm \kappa)^2-
2\varepsilon(\gamma_1\pm \kappa)\Bigr\}^{1/2}  
\label{omegapm}
\end{equation} 
in units of $\sqrt{Z_{\varepsilon}}~r_1 J_{SE}$.  
It should be noted that $\omega_\pm$ recover the orbiton energy 
$\omega_{1,2}$ when $\varepsilon=0$.  
Using bare orbiton dispersions, the Hartree decoupling parameters
$\varepsilon_{s,u}$ are calculated as follows: 
\begin{eqnarray} 
\varepsilon_s &=& \frac{3}{4} \sum_{\vec k} 
\Biggl\{ \bigl(1+\gamma_1^2 +\kappa^2 \bigr)  
+ \frac{1- \bigl( \gamma_1^2- \kappa^2 \bigr)^2}{\omega_{1 \vec k} \, \omega_{2 \vec k}} \Biggr\}  
\nonumber \\ 
&&\hspace{3em}\times  
\frac{1}{\omega_{1 \vec k} + \omega_{2 \vec k}}, 
\\ 
\varepsilon_u &=& \frac{3}{4} \sum_{\vec k} 
\Bigl\{ \bigl(1-\gamma_1^2 -\kappa^2 \bigr) + \omega_{1 \vec k} \, \omega_{2 \vec k} \Bigr\}  
\frac{1}{\omega_{1 \vec k} + \omega_{2 \vec k}}.  
\end{eqnarray} 
Numerical calculation gives $\varepsilon_s=1.72$ and $\varepsilon_u=0.59$, 
reflecting that staggered fluctuations 
of densities are stronger. 
Thus we obtain $\sqrt{Z_{\varepsilon}}=1.82$ and $\varepsilon \simeq 0.18$. 
Dispersion relations of the orbital excitations, Eq.(\ref{omegapm})  
calculated using these parameters are presented in Fig.~\ref{fig:dis_gap} 
(in units of $\sqrt{Z_{\varepsilon}} ~r_1 J_{SE}$). Staggered
density fluctuations are gapped, and we are left now with true Goldstone
phase modes protected by the symmetry of interactions.  
 
\section{Why YT\lowercase{i}O${_3}$ has a ferromagnetic ground state} 
\label{sec:why} 
So far, we discussed $t_{2g}$ orbital physics on an ideal 
cubic lattice assuming a spin saturated ferromagnetic state. 
In the remainder of the paper, we apply the theory to the ferromagnetic
state of Mott insulator YTiO$_3$. This requires some modifications of the
theory implementing a specific feature of this material. On empirical
grounds, it is well documented that Ti-O-Ti bond angle is an
important parameter controlling magnetic properties of titanates 
$R$TiO$_3$.\cite{KAT97} The bond angle $\theta$ gradually decreases 
from $\sim 157$~deg in LaTiO$_3$ to $\sim 142$~deg in YTiO$_3$, 
due to lanthanum contraction effect that results in deviations of
the lattice from an ideal perovskite structure.
It is quite remarkable that such a small variation of the 
bond angle, driven by $R$-ionic size effect, affects 
the magnetic state dramatically: It changes from isotropic AF as observed in 
LaTiO$_3$ to the isotropic ferromagnetic state in a Y-based compound, indicating 
strong competition between AF and F interactions in titanates.
 
Ti-O-Ti bond distortion is important because it induces an {\it unfrustrated}
ferromagnetic interaction, changing thereby a delicate balance between
AF and Ferro couplings that dynamically coexist and compete in 
ideal $t_{2g}$ superexchange models like in Eq.(\ref{Heta0}). It was found
in Sec.~\ref{subsec:orb_quadru} that the ferromagnetic state is slightly
higher in energy that $G$-type AF one; the situation is however reversed
when the bond angle is reduced below some critical value, as we argue below.   

The bond distortion brings about the following two effects.  
(i)~Reduction of transfer intensity between nearst-neighbor (NN) $t_{2g}$ orbitals as  
$t= \Delta_{dp}^{-1} t_{dp\pi}^2 \cos \theta = t_0 \cos \theta$.\cite{KAT97}  
Here, $t_{dp \pi} (t_{dp \sigma})$ is the transfer 
between Ti $3d$ and O $2p$ orbitals  
on the $\pi (\sigma)$ bond, and $\Delta_{dp}$ is the level difference  
between Ti $3d$ and O $2p$ states.  
Superexchange energy scale is then reduced 
as $J_{SE}=J_{SE}^{(0)} \cos^2 \theta$ with $J_{SE}^{(0)}=4t_0^2/U$.  
(ii)~Generation of transfer intensity between NN $t_{2g}$ and $e_g$ orbitals,  
$t'= \Delta_{dp}^{-1} t_{dp\sigma} t_{dp\pi} \sin \theta$.  
This transfer induces an additional SE interaction.  
For $c$-bond, we consider the following transfer term:  
\begin{eqnarray} 
{H'}_t^{(c)} = t' \Bigl[  
\bigl( \alpha_{i \sigma}^\dag a_{j \sigma} + h. c. \bigr) + (i \leftrightarrow j) 
\Bigr] . 
\label{eq:Ht'} 
\end{eqnarray}  
$\alpha^\dag$ denotes the creation operator of 
electron in the $e_g$ orbital with  
$3z^2-r^2$ symmetry.  
${H'}_t$ for $a$- and $b$-bonds are given by replacing $(\alpha,a)$ 
in Eq.~(\ref{eq:Ht'}) by  
$[-(\alpha - \sqrt{3} \beta)/2,b]$ and $[-(\alpha + \sqrt{3} \beta)/2,c]$, 
respectively, where $\beta$ denotes the electron annihilation operator 
in the $e_g$ orbital with $x^2-y^2$ symmetry.  
By the second-order perturbation with respect to $H'_{t}$,  
one obtains the SE interaction between NN $t_{2g}$ electrons.  
Here, energy of the intermediate $d^2$ excited states 
with spin triplet and singlet states  
between $e_g$ and $t_{2g}$ electrons is given by  
$U-3J_H+\Delta_{cr}$ and $U-J_H+\Delta_{cr}$, respectively,  
where $\Delta_{cr}$ is a cubic crystal-field splitting 
between $e_g$ and $t_{2g}$ levels (so-called $10Dq$).  
Explicit expression for the new SE interaction is  
\begin{eqnarray} 
{H'}_{SE}^{(c)} &=& -\frac{1}{8} J_{SE} \biggl(\frac{t'}{t} \biggr)^2 \frac{U}{\widetilde U}  
\frac{1}{(1-3\tilde \eta)(1-\tilde\eta)} \nonumber \\ 
&& \times \bigl( 2-3 \tilde \eta + 4 \tilde \eta \vec S_i \cdot \vec S_j \bigr)  
(n_{ia} + n_{ja}), 
\label{eq:SE'} 
\end{eqnarray} 
where $\widetilde U = U+\Delta_{cr}$ and $\tilde \eta = J_H/\widetilde U$.  
${H'}_{SE}$ for $a$($b$) bonds are given by replacing $n_a$ 
in Eq.~(\ref{eq:SE'}) by $n_b$ ($n_c$).  
It is stressed that this SE interaction is of the ferromagnetic sign,  
because $t_{2g}$ and $e_g$ orbitals are of the different symmetry, and  
the Hund coupling between them favors spin triplet state.  
As Ti-O-Ti bond angles in $a$, $b$, and $c$-directions are almost 
the same,\cite{KAT97} this interaction supports ferromagnetism 
equally in all three directions. Here we differ from 
Ref.~\onlinecite{MOC01}, which considers $t_{2g}-e_g$ hopping channel
along the $c$-axis only. 
 
Either in the orbital-ordered states [I (a) and I (b)] with  
$\psi_{\tilde c} = \frac{1}{\sqrt{3}}(d_{yz}+d_{xz}+d_{xy})$, or  
in the orbital-liquid one, average occupation number of each $t_{2g}$ orbital 
is given by $n_a=n_b=n_c=1/3$.  
Thus, the spin interaction (\ref{eq:SE'}) in these orbital states becomes  
$H_{spin}' =- J' \sum_{\langle ij \rangle} \vec S_i \cdot \vec S_j$  
with 
\begin{equation} 
J'= \frac{1}{3} J_{SE} \biggl(\frac{t'}{t} \biggr)^2  
\biggl(\frac{U}{\widetilde U} \biggr)^2 \eta .  
\label{eq:j'}
\end{equation} 
One should notice that $J'$ is proportional to $\sin^2 \theta$ (via $t'$), 
and contains also the small number $\eta=J_H/U$.  
This is because $J'$ is caused by the Hund coupling between 
$t_{2g}$ and $e_g$ electrons in the virtually excited $d^2 (e_g^1 t_{2g}^1)$ 
state, which is evoked in SE process only in the presence
of Ti-O-Ti bond angle distortion.  
\begin{figure}[t] 
\epsfxsize=0.75\columnwidth 
\centerline{\epsffile{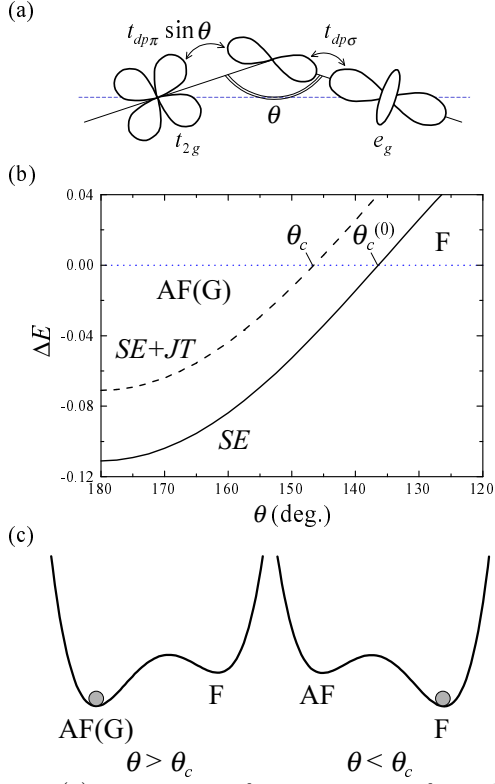}} 
\caption{(a) $t_{2g}$-$e_g$ transfer originating from Ti-O-Ti bond 
distortion. Ti-O-Ti bond angle is denoted by $\theta$.  
(b) Energy difference $\Delta E$ (in units of $J_{SE}^{(0)}$)  
between spin-AF(G)/orbital-liquid   
and spin-F/orbital-ordered states as function of $\theta$.  
The solid line is a result of purely electronic, SE interactions. 
The broken line includes small JT energy gain 
$\delta E_{JT}$ in the orbitally ordered ferromagnetic state.  
Parameters are $\eta=0.12$, $U/\Delta_{cr}=2.5$,  
$t_{dp \sigma}/t_{dp \pi}=2$, and $\delta E_{JT}/J_{SE}^{(0)}= -0.04$.  
(c) Schematic energy diagrams at $\theta > \theta_c$ (left)  
and at $\theta < \theta_c$ (right).}  
\label{fig:newSE} 
\end{figure} 

The energy difference between AF and ferromagnetic phases stemming 
from $H_{spin}' $ is given by $\Delta E'_{SE}=\frac{1}{2}J'$,  
while that from $H_{SE}$ is given by  
$\Delta E_{SE}= (-0.33\frac{r_1+r_2}{2} + 0.214 r_1) J_{SE}$ 
(Sec.~\ref{subsec:orb_quadru}). 
The total SE energy difference between AF and 
ferromagnetic phases is then estimated as  
\begin{equation} 
\Delta E =\Biggl[ -0.33\frac{r_1+r_2}{2} + 0.214 r_1 
+\frac{1}{2} \biggl(\frac{t'}{t} \biggr)^2  
\biggl(\frac{U}{\widetilde U} \biggr)^2 \eta \Biggr] J_{SE}.  
\end{equation} 
Thus, $\Delta E$ obtains the following $\theta$ dependence 
(at representative value $\eta=0.12$ for Hund's coupling parameter):  
\begin{equation} 
\Delta E =\Biggl[ -0.111 \cos^2 \theta +
\frac{1}{2} \biggl(\frac{t_{dp\sigma}}{t_{dp\pi}} \biggr)^2  
\biggl(\frac{U}{\widetilde U} \biggr)^2 \eta \sin^2 \theta \Biggr] 
J_{SE}^{(0)} . 
\end{equation} 
With the realistic parameters (typically, $\Delta_{cr}$ is about 2 eV)  
$U/\Delta_{cr}=2.5$ and $t_{dp\sigma}/t_{dp\pi}=2$, the 
transition from AF orbital-liquid phase to ferromagnetic orbital-ordered one occurs at  
the critical angle $\theta_c^{(0)}=136$~deg in the SE model.  
This angle is slightly smaller than that observed in YTiO$_3$ 
($\theta \sim 142$~deg). Further, orbital-ordered state should be favored
over the AF orbital-liquid state by orbital-lattice coupling.   
We simulate this by adding JT energy gain $\delta E_{JT} (<0)$ 
to the energy of the ferromagnetic orbital-ordered state.  
As shown in Fig.~\ref{fig:newSE}~(b), this increases $\theta_c$.
A value of $\delta E_{JT}$, which is required to obtain a realistic
value $\theta_c=146$~deg for titanates is small 
$(-0.04 J_{SE}^{(0)})$, so  
it might be hard to observe the associated $D_{3d}$-type distortion.  
 
\section{Orbital gap in YTiO$_3$} 
\label{sec:orbgap}
{\em \underline{Effect of TiO$_6$ tilting}}:  
In addition to the finite transfer between NN $t_{2g}$-$e_g$ orbitals,  
octahedron tilting 
changes also the symmetry of NN $t_{2g}$-$t_{2g}$ hopping matrix, 
making possible finite electron transfer between 
the NN orbitals with different symmetry.\cite{ISH02} 
We show now that such a hopping leads to an important modification of the 
orbital excitation spectrum, removing gapless Goldstone modes. 
Taking into account nondiagonal hoppings between orbitals active on a given
direction 
\begin{eqnarray} 
t''\Bigl[ \bigl(\alpha_i^\dag \beta_j + \beta_i^\dag \alpha_j \bigr)  
+ (i \leftrightarrow j) \Bigr]^{(\gamma)}
\label{eq:Ht''} 
\end{eqnarray} 
up to second order in small ratio $t''/t$, we obtain   
the following correction to the interaction between NN orbitals in
a spin-ferromagnetic state:  
\begin{eqnarray} 
{H''}_{SE}^{(\gamma)}&=&r_1 J_{SE} \biggl( \frac{t''}{t} \biggr)  
\bigl[\tau_{ix}^{(\gamma)}(1-n_j^{(\gamma)}) + 
\tau_{jx}^{(\gamma)}(1-n_i^{(\gamma)})\bigr] 
\nonumber \\ 
&+&\frac{1}{2}r_1 J_{SE}\biggl( \frac{t''}{t} \biggr)^2 
\bigl[B_{ij}^{(\gamma)}-n_i^{(\gamma)}n_j^{(\gamma)}\bigr].
\end{eqnarray} 
Here $B_{ij}^{(\gamma)}$ is given by either
of Eqs.~(\ref{eq:B_ab}),(\ref{eq:B_l}), and (\ref{eq:B_tau}). 
A crucial point is that this operator {\it violates} the orbital ``color''
conservation rule even in the fully spin polarized state. Therefore, uniform
phase rotations (separately on each orbital flavor) are not longer possible,
hence the relative phases will be fixed and orbital gap will be generated.

Let us apply a radial gauge description, and focus on phase fluctuations, 
as the amplitude fluctuations have a large gap anyhow.  
In a local coordinates [defined by Eq.(\ref{xyz})], we find 
\begin{eqnarray} 
{H''}_{\theta \varphi}^{(\gamma)} = \frac{1}{2} r_1 J_{SE} \rho_0^2 
\biggl( \frac{t''}{t} \biggr)^2 \sum_{\langle ij \rangle}
\Bigl( \phi_i^{(\gamma)}+\phi_j^{(\gamma)} \Bigr)^2 .  
\end{eqnarray} 
Here, we expanded the function 
$\cos \bigl(\phi_i^{(\gamma)}+\phi_j^{(\gamma)}\bigr)$, which
enters in $B_{ij}^{(\gamma)}$, about $\theta,\varphi=0$ (this corresponds 
to the quadrupole ordered state, which is in fact 
favored by $B_{ij}^{(\gamma)}$ term).

{\em \underline{Effect of trigonal TiO$_6$ distortion}}:  
Our orbital state would be supported also by a trigonal ($D_{3d}$) 
distortion of TiO$_6$. When this distortion is treated as a static one,  
an orbital feels the following potential: $H_{JT} = - 2 |E_{JT}| Q_{iz}$,  
where $E_{JT} < 0$ represents the total JT energy gain.  
In a radial gauge, we obtain the following phase Hamiltonian  
\begin{equation} 
H_{\theta \varphi}^{JT} = \frac{9}{2} \rho_0 |E_{JT}| 
\sum_{i}\bigl( \theta_i^2 + \varphi_i^2 \bigr).  
\end{equation} 
 
Summing up all contributions,  
$H_{\theta \varphi}$ [Eq.(\ref{eq:Htp})], ${H''}_{\theta \varphi}$, 
and $H_{\theta \varphi}^{JT}$, one obtains the following phase Hamiltonian:  
\begin{eqnarray} 
H_{\theta \varphi}^{tot}  
&=& H_{\theta \varphi} + {H''}_{\theta \varphi} + H_{\theta \varphi}^{JT} \nonumber \\ 
&=& Z_f \sum_{\vec q}\bigl\{\frac{1}{2}\tilde a_\theta |\theta_{\vec q}|^2  
+ \frac{1}{2}\tilde a_\varphi |\varphi_{\vec q}|^2 - 
\gamma_2^f \varphi_{\vec q} \,\, \theta_{-\vec q}\bigr\}, 
\label{ham_Zf}
\end{eqnarray} 
where $Z_f=(1+(\frac{t''}{t})^2 + 3 |E_{JT}| )$ is an overall factor.  
Here $\tilde a_\theta = 1- (\gamma_1^f + \gamma_3^f)$ and 
$\tilde a_\varphi = 1 - (\gamma_1^f - \gamma_3^f)$, with modified form factors
$\gamma_n^f = (1-2f) \gamma_n$ $(n = 1,2,3)$ where
$f = \{(\frac{t''}{t})^2+ \frac{3}{2} |E_{JT}| \}/Z_f$.

Using now Eqs.~(\ref{ham_Zf}) and (\ref{ham_Z}) for
$H_{\theta \varphi}$ and $H_{\lambda r}$, respectively, we obtain from  
$L_{orb}$ [Eq.~(\ref{eq:Lorb})] 
the following excitation spectrum:  
\begin{eqnarray} 
\omega_\pm (\vec k) = 
\Bigl\{1 - (1-2\varepsilon)(1 &-& 2f) (\gamma_1 \pm \kappa)^2  \nonumber \\
 &-& 2 (\varepsilon -f) (\gamma_1 \pm \kappa) \Bigr\}^{1/2}.
\label{omegafinal}
\end{eqnarray} 
This is given in units of $W_{orb}$, which is defined as follows: 
\begin{equation} 
W_{orb}=\sqrt{Z_\varepsilon Z_f}\,\, r_1 J_{SE}.
\label{W_orb} 
\end{equation} 
$W_{orb}$ represents the overall energy scale for orbital fluctuations 
in the problem. Excitation gaps at $(0,0,0)$ and $(\pi,\pi,\pi)$ are given by  
$2 \sqrt{f(1-\varepsilon)} \, W_{orb}$ and  
$2 \sqrt{\varepsilon(1-f)} \, W_{orb}$, respectively.  
Taking $t''/t=0.2$ and $|E_{JT}|=0.04 (r_1 J_{SE})$ as a
representative values, 
we obtain $f= 0.086$. With $\varepsilon = 0.18$ and 
$\sqrt{Z_\varepsilon}=1.82$ 
estimated above (Sec.~\ref{subsec:phases}), one obtains 
$W_{orb} \simeq 1.96~r_1 J_{SE}$, and the lowest gap about 
$0.53 W_{orb}$ is then expected at $(0,0,0)$ point. 
Thus, we expect that the orbital excitations in the modified model 
for YTiO$_3$ cover the energy window from $\sim r_1 J_{SE}$ to 
$\sim 2~r_1 J_{SE}$.
 
\section{Effective spin Hamiltonian} 
\label{sec:effective} 
The spin wave spectrum in YTiO$_3$ shows the ``cubic symmetry'' of
the Heisenberg spin couplings: $J_a\simeq J_b\simeq J_c$.\cite{ULR02}  
The magnon gap was found to be very small, 
almost two orders of magnitude smaller that the magnon bandwidth ($\sim 20$ meV).
It has been noticed that such an apparent simplicity of spin excitations,
showing high isotropy in both real and spin spaces, is remarkable and
puts strong constraints on possible orbital orderings.
Spin wave excitations are examined in this section. Being a test case 
for the above theory for orbitals in YTiO$_3$, a comparison with 
experiment gives also an opportunity to estimate SE energy scale $J_{SE}$
in the problem. To derive an effective Hamiltonian
describing magnon excitations, we assume that orbital-spin separation
occurs at low energies. This is justified when the orbital gap
induced by Ti-O-Ti bond distortions (see Sec.~\ref{sec:orbgap}) is 
larger than magnon energy. Dynamical coupling between the spin and orbital
degrees of freedom via fluctuations of superexchange bonds and 
also via on-site spin-orbital interaction $H_{so}$ is then considered
as a high-energy process, leading to an effective spin Hamiltonian.
The parameters of such a Hamiltonian are obtained by integrating out
high-energy orbital fluctuations. 
 
\subsection{Isotropic spin exchange} 
\label{subsec:iso} 
 
We start with estimation of coupling constant $J$ in the isotropic 
spin exchange term, $J(\vec S_i \cdot \vec S_j)$. 
As a first step, let us consider mean-field approximation, in which 
the spin exchange is given by an expectation value of the 
orbital operator in Eq.~(\ref{eq:Jgamma}). Neglecting a small term 
$\frac{r_2-r_3}{3r_1} \langle B_{ij}^{(\gamma)} \rangle$ and noticing that  
$\langle A_{ij}^{(\gamma)} \rangle = \frac{2}{3} E_0$,  
one obtains  
\begin{eqnarray} 
J_0 = \biggl\{ -\frac{2}{3} \eta r_2  
- \frac{2}{3} (1-\eta r_2) |E_0| \biggr\} r_1 J_{SE} .  
\label{eq:j0_av} 
\end{eqnarray} 
The first term ($\sim \eta$) is driven by a conventional Hund's coupling, 
while the second one originates from orbital  
singlet correlations in the ground state.  
For $\eta =0.12$, these two (classical and quantum) contributions  
are of the same order and give  
together $J_0 \simeq -0.214(r_1 J_{SE})$.  

\begin{figure} 
\epsfxsize=0.8\columnwidth 
\centerline{\epsffile{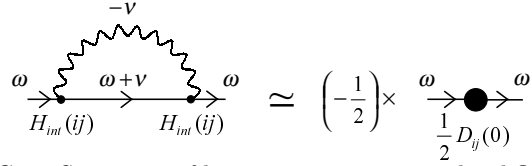}} 
\caption{Scattering of low energy magnons on local fluctuations 
of the spin exchange integral, $\delta \hat J_{ij}$.  
Its Green's function (wavy line) is taken at zero frequency 
(adiabatic approximation, see text). This results in an additional 
effective spin exchange constant (denoted by a filled circle 
on right hand side),  
$\delta J_{eff} = \frac{1}{2}D_{ij}(0)$ for low energy magnons.}  
\label{fig:scatter} 
\end{figure}

However, the actual value of $J$ measured experimentally from magnon spectra 
could in fact be strongly reduced from $J_0$ in Eq.~(\ref{eq:j0_av}) 
due to a fluctuation effects.  
Indeed, AF and F states are strongly competing in $t_{2g}$ systems,  
and large-scale orbital fluctuations are expected  
to bring about AF spin exchange contribution.  
We, therefore, have to consider effects of the dynamical 
spin-orbital interaction:  
\begin{equation} 
H_{int} = \sum_{\langle ij \rangle} 
\delta (\vec S_i \cdot \vec S_j) \; \delta \hat J_{ij} \;.  
\label{eq:mag_orb} 
\end{equation} 
In a ferromagnetic state  
$\delta (\vec S_i \cdot \vec S_j) \simeq 
-\frac{1}{2}(s_i^\dag -s_j^\dag) (s_i -s_j)$,  
with $s_i^\dag $ being a magnon creation operator.  
Neglecting small $\frac{r_1-r_2}{2r_1}$ and $\frac{r_2-r_3}{3r_1}$ terms 
in Eq.(\ref{eq:Jgamma}), exchange integral fluctuations are given by   
$\delta \hat J_{ij} \simeq \delta A_{ij}^{(\gamma)}$ 
(in units of $r_1 J_{SE}$).  
As the coupling constant in Eq.~(\ref{eq:mag_orb}) is not small,  
and because of spins and orbitals may form bound states\cite{KHA00}  
in an excited AF states,  
we will discuss here only a qualitative picture.  
We introduce a correlation function  
$D_{ij}^{(\gamma)} (\tau) =  
\langle  T_\tau \delta \hat J_{ij} (0) \delta \hat J_{ij} (\tau) \rangle$  
describing fluctuations of the spin exchange integral.  
We assume that its spectral function  
$\rho(\omega) = \frac{1}{\pi} D''_{ij} (\omega + i \delta)$ is distributed  
over the characteristic energies larger than  low-energy coherent magnons observed  
in the experiment (an adiabatic approximation which is valid as far as one is  
concerned with low-energy spin excitations).  
Within this approximation and neglecting vertex corrections we may evaluate  
the magnon scattering process on a given bond as described in Fig.~\ref{fig:scatter}.  
The result implies a renormalization of the coupling constant in low energy  
spin Hamiltonian by  
$\delta J_{eff} = \frac{1}{2} D_{ij} (0) =  
\int_0^\infty \rho (\omega) \frac{d \omega}{\omega}$,  
which is of AF sign as expected.  
It is the renormalized exchange coupling $J = J_0 + \delta J_{eff}$ that determines  
magnon spectra.  
We can estimate $D_{ij} (\omega)$ by keeping in  
$\delta \hat J_{ij} \simeq \delta A_{ij}$ [Eq.~(\ref{eq:Hrot3})] 
the orbiton pair excitation terms only:  
\begin{eqnarray} 
\delta \hat J_{ij}^{(c)} (pair) = \frac{1}{3}  
(\tilde a_i^\dag \tilde a_j^\dag +\tilde b_i^\dag \tilde b_j^\dag - 
\tilde a_i^\dag \tilde b_j^\dag - \tilde b_i^\dag \tilde a_j^\dag + H. c.).  
\end{eqnarray} 
We expect that orbiton pair fluctuations are rather incoherent and local,  
and we parametrize their spectral function by a characteristic energy  
$\Omega_{pair}$, obtaining  
\begin{eqnarray} 
D_{ij} (i \nu) =  
\frac{4}{9} \frac{2 \Omega_{pair}}{\Omega_{pair}^2 + \nu^2},  
\label{eq:Dij} 
\end{eqnarray} 
thus $\delta J_{eff} = \frac{4}{9} \frac{1}{\Omega_{pair}}$. 
We notice that Eq.~(\ref{eq:Dij}) means also that  
$\langle (\delta J_{ij})^2 \rangle = \frac{4}{9}$,  
which can be simply understood as follows:  
on a given exchange bond, {\it e. g.,} along the $c$-axis, one  
may have in general nine orbital configurations.  
>From Eq.~(\ref{eq:A_tau}) one observes that these are   
the orbital singlet giving ferromagnetic spin exchange $J=-1$,  
orbital triplet states with $J=+1$,  
and five states with $n_i^{(\gamma)}n_j^{(\gamma)} =0$ giving zero $J$.  
As orbital order is weak, and because of the formation of orbital 
singlets/triplets  
at given bond necessarily frustrates other neighboring bonds,  
all these configurations will be present giving  
$\langle (\delta J_{ij})^2 \rangle \sim \frac{4}{9}$.  
It is also natural to expect that $\Omega_{pair} \sim W_{orb}$, 
with $W_{orb} \sim 2 \;(r_1 J_{SE})$ as estimated in the preceding section. 
This gives an estimation $\delta J_{eff} \simeq \frac{2}{9}$ (in units of 
$r_1 J_{SE}$), resulting finally in effective exchange coupling  
for low energy spin excitations as follows:  
\begin{eqnarray} 
J \simeq \biggl\{-\frac{2}{3}\eta r_2 -\frac{2}{3}(1-\eta r_2) |E_0| 
+\frac{2}{9}\biggr\} r_1 J_{SE}.  
\label{eq:Jnormalized} 
\end{eqnarray} 
These qualitative estimations are substantiated in 
Appendix \ref{sec:coherent}, in which we calculate magnon energy 
renormalization within a linear orbital wave theory.  
 
It is observed from Eq.~(\ref{eq:Jnormalized}) that $J$ is actually positive 
(antiferromagnetic) for realistic values of $\eta \sim 0.12-0.13$,  
in agreement with the conclusion obtained above from energy considerations:  
The ground state of the model is not ferromagnetic in an ideal cubic lattice. 
At the presence of Ti-O-Ti bond angle distortion, $J$ is however modified as follows:  
Eq.~(\ref{eq:Jnormalized}) obtains a prefactor $\cos^2 \theta$, and in addition  
a term $-J' \simeq -\frac{2}{3} \eta \sin^2 \theta$ [Eq.~(\ref{eq:j'})]  
has to be accounted for.  
As a result, a classical Hund's rule part of $J$ remains unchanged, and the net result  
\begin{eqnarray} 
J=\biggl[ -\frac{2}{3}\eta r_2 + \cos^2 \theta  
\biggl\{ \frac{2}{9} -\frac{2}{3}(1-\eta r_2) |E_0| \biggr\} 
\biggr] r_1 J_{SE}^{(0)}  
\label{eq:Jtheta} 
\end{eqnarray} 
gives a small ferromagnetic coupling $J \simeq -0.03 r_1 J_{SE}^{(0)}$ 
for YTiO$_3$ with $\theta \approx 142$~deg. 
Comparing this result (at $\eta =0.12$) with experimental one 
$J_{exp.} = -2.75$~meV, \cite{ULR02} we obtain  
the overall energy scale $r_1 J_{SE}^{(0)} \simeq 92$~meV  
and $J_{SE}^{(0)}=4t_0^2/U \simeq $ 59~meV. Bond distortion effect
reduces the energy scale to $r_1J_{SE}$  
$\sim$ 78 meV and $\sim$ 57 meV in La and Y based titanates, respectively.  
Based on the above considerations,  
we consider $J_{SE}^{(0)} \sim 60$~meV and  
$r_1 J_{SE} \sim 60$~meV as representative energy scales for YTiO$_3$.  
 
A main message of the above considerations is that the spin-exchange 
constant as seen by a coherent low-energy magnon excitations 
in YTiO$_3$ represents in fact only a small fraction of the real 
strength of dynamical spin couplings. Because the sign of $t_{2g}$-spin 
exchange is not unique, and because the orbital order is weak, 
large fluctuations of the spin couplings are present in titanates.  
 
\subsection{Anisotropic SE interaction} 
\label{subsec:ani} 
Next, we consider effects of 
a relativistic spin-orbit coupling  
\begin{eqnarray} 
H_{so} = 
\lambda \sum_i \Bigl( \vec S_i \cdot \vec l_i \Bigr).
\label{eq:Hso} 
\end{eqnarray} 
This interaction introduces anisotropy in the effective spin Hamiltonian, 
which obtains (besides the rotationally invariant Heisenberg part) an additional,
so-called 
{\em antisymmetric Dzyaloshinskii-Moriya} (DM) 
and {\em symmetric anisotropy} interactions.\cite{MOR60} The anisotropic 
interactions 
select orientation of the 
magnetization in the crystal, and lead also to magnon gap(s). 
The structure of anisotropic terms is essentially determined by orbital state 
via expectation values and dynamics of the angular momentum operator  
in Eq.(\ref{eq:Hso}). Thus, we would like to obtain spin-orbit coupling 
induced corrections to the spin Hamiltonian, and discuss their consequences 
on magnon spectra, thereby testing the proposed orbital state.  

As usual, anisotropic interactions are obtained by perturbation 
theory involving both isotropic $H_{SE}$ and $H_{so}$. 
We mostly discuss the quadrupole ordered orbital state with
condensed $\tilde c$ orbitals (see Sec.~\ref{subsec:orb_quadru}). 
In derivation of the anisotropy Hamiltonian, 
we need to keep in the superexchange operators  
$\hat J_{ij}^{(\gamma)}$ and $\hat K_{ij}^{(\gamma)}$ in
Eq.(\ref{eq:original})   
such terms that (i) operate in the $\tilde a \tilde b$ excited states and/or  
(ii) connect a ground state with excited states of orbitals. For instance, 
\begin{eqnarray} 
\hat J_{ij}^{(c)} \Rightarrow J_{SE} \frac{r_1+r_2}{6} 
&[& n_{i \tilde a} + n_{j \tilde a} + \tilde a_i^\dag \tilde a_j^\dag + \tilde a_i \tilde a_j  
- \tilde a_i^\dag \tilde b_j^\dag - \tilde a_i \tilde b_j \nonumber \\ 
&&+ (a \leftrightarrow b)].  
\end{eqnarray} 
[Here, the terms proportional to small numbers $(r_1-r_2)$ and $(r_2-r_3)$ 
are neglected].  
 
We consider nearest-neighboring sites $i$ and $j$.  
The local excitation energy to create an orbiton 
$\tilde a$ or $\tilde b$ is denoted as $\Delta_{loc}$. It is reasonable
to associate $\Delta_{loc}$ with the ``center of gravity'' of the orbiton band 
that covers the energy window from $\sim r_1 J_{SE}$ to 
$\sim 2~r_1 J_{SE}$ as obtained in Sec.~\ref{sec:orbgap}. Thus, we will
consider 
$\Delta_{loc}\sim 1.5 r_1 J_{SE}$ in our estimations, when we compare 
later on the results with experiment. 

To obtain spin anisotropy interactions, it is convenient to work again
in a rotated basis, applying transformation (\ref{R}) also for
the spins. The scalar product of NN spins in $H_{SE}$  
is then expressed as  
$\vec S_i \cdot \vec S_j = \vec {\widetilde S}_i \widetilde T^{(\gamma)} \vec {\widetilde S}_j$ where  
\begin{eqnarray} 
\widetilde T^{(a)} = \frac{1}{3}  
\left( 
\begin{array}{rrr} 
-1 &  2 & -2 \\ 
 2 & -1 & -2 \\ 
-2 & -2 & -1 
\end{array} 
\right)  
\label{eq:Ta} 
\end{eqnarray} 
and 
\begin{eqnarray} 
\widetilde T^{(b,c)} = \frac{1}{3}  
\left( 
\begin{array}{rrr} 
-(1 \mp 2s) &         -1  & 1 \pm 2s \\ 
         -1 & -(1 \pm 2s) & 1 \mp 2s \\ 
  1 \pm 2s  &   1 \mp 2s  &       -1 
\end{array} 
\right)  
\label{eq:Tbc} 
\end{eqnarray} 
for the state I~(a). [For the state I~(b), matrices $\widetilde T$ for bonds
$a$ and $b$ are equal and given by 
Eq.~(\ref{eq:Ta})].  
  
Third-order perturbation with respect to $H_{so}$ and $H_{SE}$  
gives a {\it symmetric} part of the spin anisotropy Hamiltonian 
$H_{ani}$ in a local coordinate:  
\begin{eqnarray} 
H_{ani}^{(\gamma)} =  
- \frac{1}{2} A 
\vec {\widetilde S}_i \widetilde M^{(\gamma)} \vec {\widetilde S}_j ,  
\label{eq:OurHsym} 
\end{eqnarray}  
with  
\begin{eqnarray}  
\widetilde M^{(c)} = \frac{1}{2}  
\left(  
\begin{array}{rrr} 
   2 &    1 & c-3s \\ 
   1 &    2 & c+3s \\ 
c-3s & c+3s &  -1 
\end{array} 
\right) . 
\end{eqnarray} 
Anisotropy constant $A$ in Eq.~(\ref{eq:OurHsym}) is given by the  
following expression: 
\begin{equation} 
A = \frac{4}{9} J_{SE} \frac{r_1+r_2}{2} 
\biggl(\frac{\lambda}{\Delta_{loc}}\biggr)^2. 
\label{eq:A} 
\end{equation} 
Transformation of the spin operator from a local to global coordinates  
is expressed as ${\vec S}_i = \hat R_i \hat R \vec{\widetilde S}_i$.  
The matrices $\hat R_i$ ($i=1,2,3$ and 4) transforming spin coordinates 
at four sublattices are as follows:  
\begin{eqnarray} 
\hat R_1 =  
\left( 
\begin{array}{ccc} 
 1 & 0 & 0 \\ 
 0 & 1 & 0 \\ 
 0 & 0 & 1 
\end{array} 
\right),  
\hat R_2 =  
\left( 
\begin{array}{ccc} 
-1 & 0 & 0 \\ 
 0 &-1 & 0 \\ 
 0 & 0 & 1 
\end{array} 
\right), \nonumber \\ 
\hat R_3 =  
\left( 
\begin{array}{ccc} 
-1 & 0 & 0 \\ 
 0 & 1 & 0 \\ 
 0 & 0 &-1 
\end{array} 
\right),  
\hat R_4 =  
\left( 
\begin{array}{ccc} 
 1 & 0 & 0 \\ 
 0 &-1 & 0 \\ 
 0 & 0 &-1 
\end{array} 
\right).  
\end{eqnarray} 
Using this transformation, one obtains a symmetric anisotropy Hamiltonian  
defined in the global coordinate:  
\begin{eqnarray} 
H_{ani}^{(ij)} =  
- A \vec S_i \hat M_{ij} \vec S_j ,  
\label{eq:Hani0} 
\end{eqnarray} 
where $\hat M_{ij}$ depends on the NN bond.  
For 1-3 (2-4) bonds along the $c$ axis, $\hat M_{13(24)}$ is given by  
\begin{eqnarray} 
\hat M_{13(24)} = \frac{1}{4}  
\left( 
\begin{array}{rrr} 
   3/2 &   0  & \pm 3 \\ 
    0  &  7/2 &     0 \\ 
\pm 3  &   0  &   -3/2 
\end{array} 
\right). 
\end{eqnarray} 
The interaction matrices for NN spins on $a,b$ bonds have similar
structure:
\begin{eqnarray} 
\hat M_{12(34)} = \frac{1}{4}  
\left( 
\begin{array}{rrr} 
  -3/2 & \pm 3  &  0 \\ 
\pm 3  &    3/2 &  0 \\ 
    0  &     0  & 7/2 
\end{array} 
\right), \nonumber \\ 
\hat M_{14(23)} = \frac{1}{4}  
\left( 
\begin{array}{rrr}  
 7/2 &     0  &     0 \\ 
  0  &   -3/2 & \pm 3 \\ 
  0  & \pm 3  &    3/2  
\end{array} 
\right). 
\end{eqnarray}

\begin{figure}
\epsfxsize=0.95\columnwidth
\centerline{\epsffile{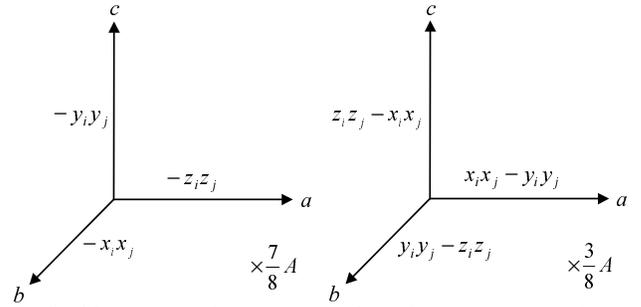}} 
\caption{Schematic representation of the structure of  
symmetric spin anisotropy interactions of $e_g$-symmetry. Interactions
along different bonds are denoted by $\alpha_i \alpha_j$, which 
should read as $S_{i \alpha} S_{j \alpha}$ times an overall
interaction constant given below each figure. For instance, 
$a$-bond interactions are $-\frac{7}{8}A S_{iz} S_{jz}$
and $\frac{3}{8}A (S_{ix} S_{jx}- S_{iy} S_{jy})$ (note 
$e_g(3z^2-r^2)$- and $e_g(x^2-y^2)$-type symmetry),
where the constant $A$ is defined by Eq.(\ref{eq:A}). 
Overall cubic symmetry of the interactions is evident 
for both contributions.
}  
\label{fig:AniEg} 
\end{figure}
\begin{figure}
\epsfxsize=0.95\columnwidth 
\centerline{\epsffile{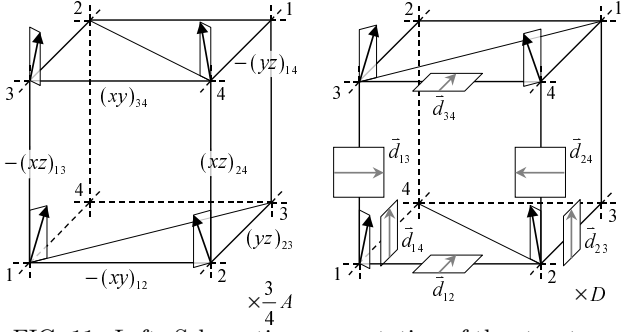}} 
\caption{Left:
Schematic representation of the structure of 
symmetric spin anisotropy interactions of $t_{2g}$-symmetry.
The notation $(\alpha \beta)_{ij}$ stands for 
$S_{i \alpha} S_{j \beta} + S_{i \beta} S_{j \alpha}$ multiplied
by interaction constant $3A/4$.  
Black arrows represent the direction of spins favorable for this interaction  
when the direction of the uniform moment is taken along [001].  
Right: Antisymmetric DM spin anisotropy interactions [Eq.(\ref{eq:Hdm})].  
Gray arrow denoted by $\vec d_{ij}$ shows the orientation of DM vectors
on different bonds. A preferred spin pattern for this interaction 
is shown by black arrows.}
\label{fig:AniT2DM} 
\end{figure}

 
Symmetric anisotropy interactions can be classified according 
to cubic invariants:  
there are terms of $e_g$ and $t_{2g}$ symmetries, generated by diagonal 
and nondiagonal elements of the matrices $\hat M_{ij}$, correspondingly.  
For convenience, we show the interactions for the state I~(a)    
in Fig.~\ref{fig:AniEg} and Fig.~\ref{fig:AniT2DM}~(a), in which bond dependence, direction of spins  
and the scale of individual anisotropy terms are shown.  
The $e_g$ symmetry anisotropy in Fig.~\ref{fig:AniEg}~(a) has been discussed in  
Ref.~\onlinecite{KHA01} under the name of ``cubic'' anisotropy in the context of magnon gap in LaTiO$_3$.  
A remarkable feature of this interaction is its intrinsic frustrations:  
namely, treated classically, it acquires a rotational symmetry in spin 
sector, resulting in an infinite degeneracy of classical states.  
An accidental pseudo-Goldstone mode, which appears in classical limit, can
acquire finite gap by quantum fluctuations only.  
In the present orbital ordered states, 
the symmetry is lowered compared with the orbital liquid state in LaTiO$_3$.  
Thus, additional terms are generated as shown in 
Figs.~\ref{fig:AniEg}~(b) and ~\ref{fig:AniT2DM}~(a).  
It is noticed that these terms have a similar frustrated nature:  
summed over all the bonds, they cancel each other exactly.  
Thus, only a small gap is expected from these interactions.
  
Physically, the structure of anisotropy interactions is determined by local
correlations of the angular momentum, and can therefore be traced back 
to noncollinear
arrangements of these correlations shown in Fig.~\ref{fig:2pattern}~(a).  
For instance, the leading, ``cubic'' term [see Fig.~\ref{fig:AniEg}~(a)]
reflects that $l_y$ components are correlated ferromagnetically along
the $c$ axis, while $l_z (l_x)$ components are parallel along $a(b)$ axes.
 
In the state I~(b), the leading anisotropy interaction 
in the $ab$-plane is given by $-S_i^z S_j^z$ and that along $c$-direction 
is obtained to be $-S_i^y S_j^y$ 
[as can easily be seen also from Fig.~\ref{fig:2pattern}~(b)].  
The number of bonds with anisotropic interaction $-S_i^z S_j^z$ is two times
larger than that with $-S_i^y S_j^y$.  
This breaks the rotational symmetry in a spin sector even in the classical limit,
and generates a large magnon gap $\Delta_{mag} = \frac{7}{4} \sqrt{2} SA$.  
 
{\it Antisymmetric DM} interaction appears in a second-order perturbation 
theory as a combined effect of $H_{SE}$ and $H_{so}$. 
The remarkable feature of the orbital state I is that leading terms in 
SE interactions, which are proportional 
to $(r_1+r_2)$ in Eq.(\ref{eq:Jgamma}), do not contribute to the DM 
interaction.  
That is because of the classical expectation value of $A_{ij}$ in 
Eqs.~(\ref{eq:A_ab}) and (\ref{eq:A_l}) vanishes in our orbital states.  
Rather much smaller Hund's coupling terms proportional 
to the small number $(r_1-r_2)$ only give rise to DM interaction.  
This feature contrasts with that in the orbital state reported in  
Refs.~\onlinecite{MIZ96,SAW97,AKI01,ITO01,NAK02},   
in which a large DM interaction is present (see Appendix~\ref{sec:AniMiz}).  
 
After somewhat tedious but straightforward calculations
one obtains the following interaction between NN sites:  
\begin{eqnarray} 
H_{DM}^{(\gamma)} =  
D \vec {\widetilde S}_i \widetilde N^{(\gamma)} \vec {\widetilde S}_j . 
\end{eqnarray} 
Here, the matrix $\widetilde N^{(c)}$ reads as  
\begin{eqnarray} 
\widetilde N^{(c)} = \frac{1}{3}  
\left( 
\begin{array}{rrr} 
2c - 2s &      1  & c + s \\ 
     1  & 2c + 2s & c - s \\ 
 c + s  &  c - s  &    -2 
\end{array} 
\right),  
\end{eqnarray} 
and the interaction constant $D$ is obtained as follows:  
\begin{equation} 
D = J_{SE} \frac{r_1-r_2}{6} \frac{\lambda}{\Delta_{loc}}.  
\label{eq:defD} 
\end{equation} 
Transforming the local spin axes to the global ones,  
one arrives at the following DM interaction:  
\begin{eqnarray} 
H_{DM}^{(ij)} =  
D \vec d_{ij} \cdot \Bigl( \vec S_i \times \vec S_j \Bigr), 
\label{eq:Hdm} 
\end{eqnarray} 
with $\vec d_{ij}=\hat \alpha_i'$.  
Here, $\hat \alpha_i$ is the unit vector parallel to one of local axes  
($x_i$, $y_i$ and $z_i$) 
which is perpendicular to the $i$-$j$ bond direction and antiparallel to  
its counterpart at site $j$ [see Fig.~\ref{fig:2pattern}~(a)].  
For example, $\vec d_{13}=(1,0,0)$, $\vec d_{12}=(0,1,0)$,  
$\vec d_{14}=(0,0,1)$, {\it etc}. For convenience, we show the DM 
interaction for state I~(a) in Fig.~\ref{fig:AniT2DM}~(b).  
  
\subsection{Spin waves: comparison with experiment} 
 
Now, we discuss the anisotropic spin interactions in the context of  
the experimental observations of magnon dispersion in YTiO$_3$.\cite{ULR02}  
 
\label{subsec:spinwave} 
{\em \underline{Cubic symmetry of the spin wave dispersion}}:
This puzzling observation is naturally explained by the present theory, as  
ferromagnetic couplings in all the states I and II are perfectly isotropic, 
$J^{(a)}=J^{(b)}=J^{(c)}$. The reason is high symmetry of the orbital
ordering patterns, as can be visualized from Fig.~\ref{fig:order}. 
It is stressed that this result is robust, ``no fine-tuning''
property of the model. 
(The isotropy is expected to be relaxed somewhat by lattice distortions.
However, effects of two types of distortions, that is,  
Ti-O-Ti bond angle distortion and elongation of TiO$_6$ octahedron,  
on the anisotropy of spin couplings are opposite and  
almost cancel each other.\cite{OKAUN}) 
 
{\em \underline{Isotropy in spin space, magnon gap}}: Magnons in 
YTiO$_3$ are almost gapless\cite{ULR02} (upper limit for the gap 
is 0.3 meV); this is a serious test for possible orbital 
orderings. We show now that high symmetry of orbital orderings in the present 
theory resolves this problem as well. The crucial point is the frustrating
nature of the anisotropic spin interactions obtained above: Even though 
anisotropic couplings on an individual bond are substantial, 
there is cancellation of classical contributions stemming from 
different bonds, and one obtains only small gaps of quantum origin. 
We illustrate this by considering first the leading ``cubic'' term. 
 
{\it Quantum magnon gap by ``cubic'' term}:  
Consider the effective spin Hamiltonian in the state I~(a) given as follows:  
\begin{eqnarray} 
H_s &=& H_{iso} + H_{ani} \nonumber \\ 
&=& - \sum_{\langle ij \rangle} \Bigl[  
J \vec S_i \cdot \vec S_j + \tilde A S_i^{(\gamma)} S_j^{(\gamma)} \Bigr] .  
\label{eq:Hs} 
\end{eqnarray}  
Here,  
$H_{iso}$ represents the isotropic SE Hamiltonian.  
Coefficient of the ``cubic'' term is $\tilde A=7/8A$ 
[see Fig.~\ref{fig:AniEg}~(a)],   
and axes in the spin space are changed 
as $S^z \rightarrow S^x$, $S^x \rightarrow S^y$  
and $S^y \rightarrow S^z$ such that $\gamma$ corresponds 
to the direction of the $i$-$j$ pair. Thus, we have $- S_i^z S_j^z$ for $c$-bonds, 
$- S_i^x S_j^x$ for $a$-bonds and $- S_i^y S_j^y$ for $b$-bonds.  
 
Because the effective spin Hamiltonian $H_s$ has a discrete (cubic) symmetry,  
magnon excitation is expected to have a gap.  
However, due to the rotational symmetry of Eq.~(\ref{eq:Hs}) in the limit 
of classical spins, linear spin-wave theory cannot provide finite gap.  
This problem is resolved by the order-from-disorder 
mechanism,\cite{TSV95} which selects a particular classical state which  
provides the largest zero-point energy when fluctuations are included. This
opens also a magnon gap.\cite{KHA01,YIL95}  
Thus, we calculate spin-wave contribution to the ground-state energy 
as a function of the angle $\theta$  
between the $c$ axis and the uniform moment.  
First, we rotate the spin quantization axes around the $b$ axis:  
\begin{eqnarray}  
S_i^x &=& c_{\theta} \widetilde S_i^x + s_{\theta} \widetilde S_i^z , \nonumber \\ 
S_i^z &=&-s_{\theta} \widetilde S_i^x + c_{\theta} \widetilde S_i^z ,  
\end{eqnarray} 
where $c_{\theta} = \cos \theta$ and $s_{\theta} = \sin \theta$.  
Second, by using Holstein-Primakoff approximation, 
we obtain the magnon dispersion which shows explicit $\theta$ dependence as  
\begin{equation} 
\omega_{\vec k} (\theta) = zSJ \sqrt{X_{\vec k}[Y_{\vec k} -a (c_z -c_x) s_{\theta}^2]} ,  
\end{equation}  
where  
$X_{\vec k} = (1 - \gamma_1) + a (1 - c_y)$, $Y_{\vec k} = (1 - \gamma_1) + a (1 - c_x)$,  
and $a = \tilde A/3J$ represents the ratio between anisotropic and 
isotropic interactions.  
Finally, by calculating the zero point magnon energy, we obtain the ground-state energy $E_0$ per site. 
In the limit of $\theta \ll 1$,  
\begin{eqnarray} 
E_0 (\theta) &=&  
- \frac{1}{2}zJS(S+1)(1 + a) + \frac{1}{2} \sum_{\vec k} \omega_{\vec k} (\theta) \nonumber \\ 
&=& -const + K_{eff} S^2 \theta^2 . 
\label{potential}
\end{eqnarray} 
Here, $K_{eff}$ represents an effective spring constant.  
In the case of $a \ll 1$ which we are interested in, the spring constant is given by  
$K_{eff} = \tilde A^2 R / zSJ$ with  
\begin{equation} 
R = 3 \sum_{\vec k} \frac{\gamma_2^2}{1 - \gamma_1} \approx 0.28 \;.  
\label{eq:R} 
\end{equation} 
The potential $E_0(\theta)$ in Eq.(\ref{potential}) can be associated with an
effective uniaxial spin anisotropy Hamiltonian  
\begin{equation} 
H_{eff}^{ani} = - K_{eff} \sum_{\langle ij \rangle_c} S_i^z S_j^z ,  
\end{equation} 
generated in the symmetry-broken phase (with spins oriented along [001]).  
Therefore, one finds a magnon gap 
$\Delta_{mag} = 2 S K_{eff} = 2 \tilde A^2 R / zJ$.  
Note that $\Delta_{mag}$ is independent of $S$ and is proportional to 
$\tilde A^2$. We confirmed that the same magnitude of the magnon gap is derived 
from $H_s$ using single mode approximation \cite{FEY72} 
(see Appendix~\ref{sec:SMA}). Thus, ``cubic'' anisotropy gives the
magnon gap $\Delta_{mag} = 2 (\frac{7}{8} A)^2 R / zJ$.   

Now, we estimate the coupling constant $A$.  
>From Eq.~(\ref{eq:A}) with $r_1 J_{SE} \sim 60$~meV and $\eta \sim 0.12$,  
one finds $A \simeq 23 (\lambda/\Delta_{loc})^2$.  
As discussed in the precedings section, we consider 
$\Delta_{loc} \sim 1.5 \, r_1 J_{SE} \sim 90$~meV. 
Using the atomic value $\lambda_{at} = 19$~meV,\cite{ABR70} 
one obtains then $A \simeq 1$~meV consistent with the experimental value
of $A$ obtained in Ref.~\onlinecite{ULR02} from magnon spectra.   
With this value of $A$, the magnon gap from the ``cubic'' term is only  
$\approx 0.03$~meV in the state I~(a). 
 
On the other hand, a large classical magnon gap 
$\Delta_{mag} = \frac{7}{4} \sqrt{2} SA \approx 1.2$~meV is obtained
in the state I~(b), as we already discussed in the preceding section.
This allows us to exclude the orbital ordering structure I~(b) 
from possible candidates for YTiO$_3$, although we do not know precisely
which kind of lattice distortions favor state I~(a) over configuration I~(b).
  
{\it Contributions from the other terms}:  
Anisotropy of $e_g(x^2-y^2)$ symmetry [see Fig.~\ref{fig:AniEg}~(b)] 
can be analyzed similarly; we find that it also supports an easy 
magnetization axis along one of the cubic axes, say [001]. Its contribution
to the magnon gap, $\Delta_{mag} = 6(\frac{3}{8}A)^2R /zJ\simeq 0.014$~meV 
is smaller than that of the ``cubic'' term, as expected.
 
Once the [001] direction is chosen as the direction of uniform moment, spin
anisotropy terms of $t_{2g}$ symmetry and DM interactions give rise 
to spin cantings as shown in Figs.~\ref{fig:AniT2DM}~(a,b).  
The canting angle $\theta$ is given by 
$\theta \approx (3/2 \sqrt{2})(A/4J) \simeq 0.1$~rad for $t_{2g}$ symmetric 
anisotropy interaction, and $\theta \approx \sqrt{2}(D/4J) \simeq 0.07$~rad 
for
the antisymmetric DM interaction. [DM interaction constant $D\approx 0.57$~meV 
is estimated from Eq.(\ref{eq:defD})]. 
These values are within the experimental canting angles 
$\simeq 0.17$ rad.\cite{ULR02} Finally, contributions of $t_{2g}$ symmetric
anisotropy and DM interactions to the magnon gap $\Delta_{mag}$ 
are estimated as $\approx 4JS\theta^2(t_{2g}) \sim 0.05$~meV 
and $\approx 4JS\theta^2(DM) \sim 0.03$~meV, correspondingly; 
these numbers are rather small again. It should be noticed, 
that a more quantitative analysis of the problem, 
in particular the precise structure of the spin canting pattern 
requires consideration of all the anisotropy terms  
on equal footing, which will be presented elsewhere. 
   
To summarize this section, we have obtained spin anisotropy interactions
induced by the spin-orbit coupling, and considered their effects on spin-wave
spectra. Because of high symmetry of the orbital ordering, particularly
in the quadrupole ordered state I~(a), magnon dispersion is found to have
cubic symmetry, magnon gap is small, and there are small cantings of spins
away from the $c$ axis. All these observations are consistent with experiment. 
    
\section{Orbital contribution to the resonant x-ray scattering} 
We turn to the discussion of further experiments which may
help to verify the proposed orbital state in YTiO$_3$. First, we consider 
the resonant x-ray scattering, 
which has proven to be a useful method in the study
of orbital order symmetry.\cite{MUR98,ISH98} The following section will be devoted
to possible ways of detecting orbital excitations.

We focus on the orbital state I~(a) [which is the most plausible candidate 
as discussed in previous sections]. While the exchange bonds in this state 
are the same (important for the isotropy of spin waves), a local symmetry 
is lower than a cubic one [see Fig.(\ref{fig:order})]. Thus, orbital order 
may induce spatial modulations of the level structure of an excited 
photoelectron in $p$-states via the so-called Coulomb mechanism.\cite{ISH98}
This may lead to additional weak reflections at orbital ordering vectors.
Predictions of our theory for such an experiment are as follows.  
Orbital order shown in Fig.~\ref{fig:order}(I)  
is identified as a three-component quadrupole ordering of $t_{2g}$-symmetry,  
$T_\alpha (\vec q)$ with $\alpha =x,y,z$.  
Each component has its own propagation vector:  
\begin{eqnarray} 
T_x = \langle l_y l_z + l_z l_y \rangle_{\vec R} = \frac{2}{3} Q e^{i \vec q_1 \cdot \vec R},  
\nonumber \\ 
T_y = \langle l_x l_z + l_z l_x \rangle_{\vec R} = \frac{2}{3} Q e^{i \vec q_2 \cdot \vec R},  
\nonumber \\ 
T_z = \langle l_x l_y + l_y l_z \rangle_{\vec R} = \frac{2}{3} Q e^{i \vec q_3 \cdot \vec R},   
\end{eqnarray} 
where $\vec q_1 = (\pi,0,\pi)$, $\vec q_2 = (\pi,\pi,0)$, and $\vec q_3 = (0,\pi,\pi)$.  
As order parameter $Q$ is strongly suppressed by quantum fluctuations  
($Q \simeq 0.19$, see Sec.~III B),  
we obtain that each component has only a small amplitude, 
giving $|T_\alpha|^2 \sim 0.016$.  
This implies that the corresponding anomalous Bragg intensity is at least 60 
times weaker compared with the classical orbital orderings.  
It might therefore be very difficult to single out this contribution.  
However, new azimuthal ($\varphi$) and scattering ($\theta_s$) 
angular dependencies of an additional intensity, which should show up 
below $T_{orb}$, may help to identify order symmetry.  
For ($\pi,\pi,0$) [(100) in orthorhombic notations] scattering  
these dependences are obtained as follows  
\begin{equation} 
I_{\sigma \sigma'}(\pi,\pi,0) \propto \sin^2 2 \varphi 
\end{equation} 
for $\sigma$-$\sigma'$ polarization  
(see for notations Ref.~\onlinecite{MUR98}), and  
\begin{equation} 
I_{\sigma \pi'}(\pi,\pi,0) \propto ( \cos 2 \varphi \sin \theta_s + \sin \varphi \cos \theta_s)^2 
\end{equation} 
for $\sigma$-$\pi'$ polarization.  
(Azimuthal angle $\varphi=0$ corresponds to the configuration in which  
the diffraction plane is parallel to the $c$ axis.)  
Scattering intensities at $\vec q_1$ and $\vec q_3$ 
(which are contributed by $T_x$ and $T_z$ components, respectively)  
can be obtained from symmetry considerations.  

It is still a controversial issue whether the resonant x-ray  scattering, 
observed in YTiO$_3$,\cite{NAK02} is related to orbital order or lattice 
distortions (see Refs.\onlinecite{TAK01,NAK02,ISH02}). Either way,
we expect that the orbital order contribution, {\it if present}, must be 
temperature dependent reflecting orbital order/disorder transition, 
as in the case of manganites with strong orbital order.
Therefore, a careful analysis of the $T$-dependence of reflections
at orbital ordering vectors is desirable. 

Above discussion brings us to the problem of the orbital ordering temperature 
in YTiO$_3$.
Thus far, there are no reports on the orbital ordering temperature in 
YTiO$_3$ (weak structural change at spin ordering temperature \cite{NAK02}
is only an indirect indication). In our SE-model picture,
we expect that this transition should occur at low temperatures only, and
we suspect in fact that orbitals in YTiO$_3$ do order at ferromagnetic
spin transition $T_C$. This is because of strong spin/orbital coupling, and
also because of the frustrated nature of orbital-only model itself, which, 
as we have shown, may develop long-range order on a cubic lattice 
at zero temperature only. Lattice distortions that open a finite
orbital gap allow finite-temperature transition, 
but this cannot occur much above the ferromagnetic transition, 
because of strong disorder introduced by spin fluctuations in the
paramagnetic phase. In other words,     
orbital order and isotropic spin ferromagnetism  
are intimately connected, supporting each other.  
Physically, this implies that short-range ferromagnetic correlations are of  
vital importance for orbital ordering, and vice versa.  

A quantitative description of the finite-temperature behavior of a realistic
spin-orbital model is 
complicated. We may give only very rough estimation for the orbital
ordering temperature based on the mean-field picture. As we are going to ignore 
fluctuations completely, this estimation 
should be regarded as an upper limit, which we would like to know. 
To this end, we consider a spin paramagnetic phase and set  
$\langle \vec S_i \cdot \vec S_j \rangle=0$ in Eq.~(\ref{eq:original}),  
neglect in the orbital interactions $A_{ij}^{(\gamma)}$ 
in Eq.~(\ref{eq:Hrot3}) all the terms  
except those which contain an emerging quadrupole order parameter.  
This leads to  
\begin{eqnarray} 
H_{orb} = - \frac{1}{3z} \sum_{\langle ij \rangle} \hat Q_{iz} \hat Q_{jz} \nonumber \\ 
\Rightarrow - \frac{1}{3} \langle \hat Q \rangle \hat Q_{iz}.  
\label{eq:HorbMF} 
\end{eqnarray} 
From Eq.~(\ref{eq:HorbMF}) we obtain $T_{orb}=\frac{1}{6}$ (in units of $r_1 J_{SE}$),  
which, including the Ti-O-Ti bond angle ($\theta \simeq 142$~deg)  
correction for YTiO$_3$, reads as  
\begin{eqnarray} 
T_{orb}=\frac{1}{6}\cos^2 \theta \Bigl(r_1 J_{SE}^{(0)}\Bigr)  
\simeq 0.1 \Bigl(r_1 J_{SE}^{(0)}\Bigr). 
\label{eq:Torb} 
\end{eqnarray} 
On the other hand, spin ordering (mean-field) temperature  
is $T_C = \frac{3}{2} |J|$,  
with $J \simeq -0.03 (r_1 J_{SE}^{(0)})$ given in Eq.~(\ref{eq:Jtheta}).  
Both $T_{orb}$ and $T_C$ should, of course, be reduced by fluctuations  
(indeed, with $J_{exp}=-2.75$~meV,\cite{ULR02}  
one obtains mean-field $T_C \simeq 48$~K instead of observed 27~K),  
so it makes more sense to consider their ratio, which is  
\begin{eqnarray} 
T_{orb}/T_C \simeq 2.2 \;. 
\label{eq:Tratio} 
\end{eqnarray} 
For $T_C \sim 25-30$~K (which is sample dependent), 
this gives an upper estimation $T_{orb} \sim 55-70$~K. 
We would like to think that local orbital order,   
accompanied by short-range ferromagnetic correlations, starts to develop at
these temperatures.  
In fact, the presence of such a correlations in YTiO$_3$ below $\sim 50-60$~K 
has been reported from several experiments:  
(i) sharp drop in NMR relaxation rate, which has been speculated  
in terms of orbital ordering,\cite{FUR97}  
(ii) spin-resonance line shape changes from typical paramagnetic 
spectra to the ferromagnetic one,\cite{OKU98}  
(iii) weak quasielastic magnetic scattering is observed 
above $T_C$.\cite{ULR02}

\section{Angular momentum fluctuations: Dynamical magnetic susceptibility} 
\label{sec:susceptibility} 

In this section, we would like to calculate orbital contribution to the
inelastic neutron-scattering intensity. The point is that $t_{2g}$ orbitals
are magnetically active, as their angular momentum may directly couple to
the neutrons. Of course, there is a contribution also in nonmagnetic channels:
The orbital quadrupole moment is coupled to the phonons, and hence 
single or double orbiton (depending on the structure of this coupling)  
may be excited by neutrons indirectly via lattice vibrations. We focus
on the magnetic scattering, and calculate orbital angular momentum dynamical
susceptibility. If the $t_{2g}$ orbital level is split up by strong lattice 
distortions, one would expect just a local, crystal-field transitions. 
In SE-driven orbital picture, advocated in this paper, angular momentum 
fluctuations are however of the collective nature. Thus, we expect momentum
selected (though strongly damped) transitions, forming broad bands.
 
\subsection{Quadrupole order} 
Consider first 
local angular momentum susceptibility $\chi_{loc} (\omega)$ in the
quadrupole ordered state.  
It is defined as 
\begin{equation} 
\chi_{loc}(\omega) = \Big\langle \vec l_i \cdot \vec l_i \Big\rangle,  
\end{equation} 
and its imaginary part describes the spectral shape of the
momentum-integrated inelastic neutron-scattering cross section.  
 
\begin{figure} 
\epsfxsize=0.8\columnwidth 
\centerline{\epsffile{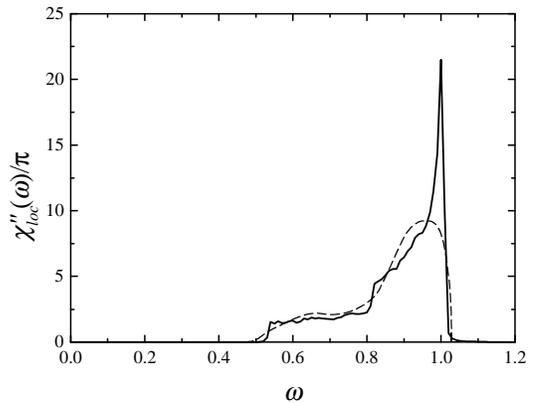}} 
\caption{Solid line:
Imaginary part of the local angular momentum susceptibility 
$\chi_{loc}'' (\omega)$ [Eq.(\ref{chilocal})] in quadrupole
ordered state of orbitals. 
The energy $\omega$ is given in units of $W_{orb}$ defined 
by Eq.(\ref{W_orb}). 
A finite gap for the orbital waves stemming from symmetry breaking
interactions is taken into account according to Eq.~(\ref{omegafinal})
with parameters $f=0.086$ and $\varepsilon = 0.18$ (this gives
$W_{orb}\simeq 2r_1J_{SE}$). 
The sharp peak structure is expected to be smoothed by damping effects 
(not accounted in the present study), as indicated by the broken line.}  
\label{fig:chi} 
\end{figure}

In a linear orbital wave approximation, 
the imaginary part of $\chi_{loc} (\omega)$ at $\omega > 0$ is given by  
\begin{equation} 
\chi''_{loc}(\omega) = \pi \sum_{\vec k}  
\biggl[ \frac{1}{\omega_{1 \vec k}} \delta(\omega - \omega_{1 \vec k})  
+ \frac{1}{\omega_{2 \vec k}} \delta(\omega - \omega_{2 \vec k}) \biggr]. 
\label{chilocal}
\end{equation} 
In order to account for a finite gap induced in the orbital sector 
by lattice distortions (see Sec.\ref{sec:orbgap}), we use hereafter
Eq.~(\ref{omegafinal}) for the orbital excitation spectrum.
The numerical result for $\chi''_{loc}(\omega)$ is presented in 
Fig.~\ref{fig:chi}. The sharp structure about $W_{orb}$ (taken as an
energy scale in the figure) is related to the orbiton band-edge effects,
which should go away when damping effects are properly taken into
account. 
 
We turn now to the momentum dependence of dynamical susceptibility,  
$\chi (\vec q, \omega) = \Big\langle \vec l_{\vec q} \cdot \vec l_{-\vec q} \Big\rangle_\omega $. 
This quantity determines a dynamical structure factor, which at $T=0$ is 
given by 
\begin{equation} 
S (\vec q, \omega) = \frac{1}{\pi} {\rm Im} \chi (\vec q, \omega) .  
\label{eq:sqomega} 
\end{equation} 
A noncollinear, four sublattice orbital order leads to the following 
structure for  
$\chi (\vec q, \omega)$: 
\begin{eqnarray} 
\chi (\vec q, \omega) &=& 
\frac{2}{3} 
\bigl[ \chi_1(\vec q + \vec q_1) + \chi_1(\vec q + \vec q_2) + \chi_1(\vec q + \vec q_3) \bigr]_\omega 
\nonumber \\ 
&& +\frac{1}{\sqrt 3} 
\bigl[\chi_2(\vec q + \vec q_2) - \chi_2(\vec q + \vec q_1) \bigr]_\omega 
\label{eq:chi} \\ 
&& +\frac{1}{3} 
\bigl[2 \chi_3(\vec q + \vec q_3) - \chi_3(\vec q + \vec q_1) - \chi_3(\vec q + \vec q_2) \bigr]_\omega .  
\nonumber 
\end{eqnarray} 
Here, the orbital ordering vectors 
\begin{eqnarray} 
\vec q_1 &=& (\pi,   0, \pi), \nonumber \\ 
\vec q_2 &=& (\pi, \pi,   0), \nonumber \\ 
\vec q_3 &=& (  0, \pi, \pi)  
\label{eq:qI} 
\end{eqnarray} 
for the state (a) and  
\begin{eqnarray} 
\vec q_1 &=& (\pi, \pi, \pi), \nonumber \\ 
\vec q_2 &=& (\pi, \pi,   0), \nonumber \\ 
\vec q_3 &=& (  0,   0, \pi)  
\label{eq:qII} 
\end{eqnarray} 
for the state (b). The susceptibilities   
\begin{eqnarray} 
\chi_1 (\vec q, \omega) =  
\frac{1}{2} \Big\langle \tilde l_{\vec q}^x \tilde l_{- \vec q}^x  
+ \tilde l_{\vec q}^y \tilde l_{- \vec q}^y \Big\rangle_\omega , \nonumber \\ 
\chi_2 (\vec q, \omega) = \frac{1}{2} \Big\langle \tilde l_{\vec q}^y \tilde l_{- \vec q}^y  
- \tilde l_{\vec q}^x \tilde l_{- \vec q}^x \Big\rangle_\omega , \nonumber \\ 
\chi_3 (\vec q, \omega) = \frac{1}{2} \Big\langle \tilde l_{\vec q}^x \tilde l_{- \vec q}^y  
+ \tilde l_{\vec q}^y \tilde l_{- \vec q}^x \Big\rangle_\omega   
\label{eq:chi123} 
\end{eqnarray} 
are defined in a rotated basis given by the transformation in Eq.~(\ref{R}).
We calculate these susceptibilities in a linear orbital-wave approximation.   
The imaginary part of $\chi_\alpha (\vec q, \omega)$ at $\omega > 0$ 
is obtained as 
\begin{eqnarray} 
\frac{1}{\pi} \chi''_\alpha (\vec q, \omega) &=&  
A_\alpha  \frac{1 + \gamma_1 + \kappa}{\omega_{1 \vec q}} \,  
\delta(\omega - \omega_{1 \vec q}) \nonumber \\ 
&&+ B_\alpha \frac{1 + \gamma_1 - \kappa}{\omega_{2 \vec q}} \,  
\delta(\omega - \omega_{2 \vec q}) ,  
\end{eqnarray} 
where $A_1=B_1=1/2$, $A_2=-B_2=\gamma_2 / 2 \kappa$ and 
$A_3=-B_3=\gamma_3 / 2 \kappa$.  
 
\begin{figure} 
\epsfxsize=0.8\columnwidth 
\centerline{\epsffile{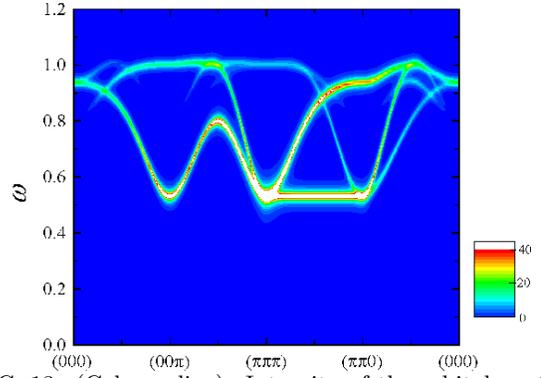}} 
\caption{(Color online). Intensity of the orbital contribution 
to the magnetic structure factor $S (\vec q, \omega)$  
[Eq.(\ref{eq:sqomega})] in the quadrupole ordered state I(a).  
Energy $\omega$ is given in units of $W_{orb}$ defined by Eq.(\ref{W_orb}). 
A finite orbital excitation gap due to symmetry breaking terms (see for
details Sec.\ref{sec:orbgap}) is taken into account according to  
Eq.~(\ref{omegafinal}) with parameters $f=0.086$ and $\varepsilon = 0.18$.}  
\label{fig:sqw} 
\end{figure}

Numerical results for $S(\vec q, \omega)$ in the state I (a) 
are shown in Fig.~\ref{fig:sqw}.  
An intensive hot spot at momentum $\vec q=(\pi,\pi,\pi)$ at energies about
orbiton gap, and flat dispersions at $(\pi,\pi,q_z)$ direction are noticed.
The rather complicated multiband structure has its origin in noncollinear 
nature of the underlying orbital ordering, characterized by a several
ordering vectors. [In the state I~(b), which has different ordering vectors 
$\vec q_i$, $S(\vec q, \omega)$ shows different energy-momentum structure   
(not shown)]. The energy scale for orbital fluctuations 
($\sim W_{orb} \sim 2r_1J_{SE}$)
is much larger than magnon energies. This is because of strong cancellation of
ferromagnetic and AF contributions to the spin-exchange integral $J$ (see Sec.
\ref{subsec:iso}), resulting in rather small magnon bandwidth 
(which is only a fraction of $r_1J_{SE}$). 
Therefore, magnon excitations are expected to be 
well defined, since they are located within the orbital gap. 
As for the high-energy orbital excitations, we expect strong damping 
effects stemming from nonlinear couplings between orbital waves 
themselves, and also from the dynamical coupling between spin and 
orbital fluctuations. These effects should in fact relax momentum resolution
and smooth away sharp structures obtained in Fig.~\ref{fig:sqw} by using
undamped orbital waves.
 
\subsection{Orbital magnetic state} 
For completeness, we also give equations for the magnetic response of
the magnetically ordered state of orbitals.
In contrast to the quadrupole ordering, this state gives rise 
to static Bragg peaks of orbital origin. These peaks are located at 
orbital ordering vectors $\vec q_i$ [Eqs.(\ref{eq:qI}) and (\ref{eq:qII})]:  
\begin{equation} 
\Big\langle \vec l_{\vec q} \cdot \vec l_{-\vec q} \Big\rangle =  
\frac{1}{3} \Big\langle m_l \Big\rangle^2  
\bigl[ \delta(\vec q - \vec q_1) + \delta(\vec q - \vec q_2) + \delta(\vec q - \vec q_3) \bigr] ,  
\end{equation} 
and their intensity is determined by the orbital magnetic order parameter
$m_l=0.19$ [Eq.(\ref{ml})].

The dynamical susceptibility 
$\chi (\vec q, \omega)$ is given by the same form as Eq.~(\ref{eq:chi}) with  
$\chi_{1, 2, 3} (\vec q, \omega)$ having the same definitions given 
by Eqs.~(\ref{eq:chi123}). In the orbital magnetic state, 
these susceptibilities 
are obtained as follows:
\begin{equation} 
\chi''_1 (\vec q, \omega) = \frac{\pi}{4} \biggl[ 
\frac{1}{\omega_{1 \vec q}} \delta(\omega - \omega_{1 \vec q})  
+ \frac{1}{\omega_{2 \vec q}} \delta(\omega - \omega_{2 \vec q}) \biggr]    
\end{equation}
and
\begin{eqnarray} 
\chi''_{2,3} (\vec q, \omega)\!= 
\!\frac{\pi\gamma_{2,3}}{4 \kappa} \biggl[\frac{\gamma_1\!+
\!\kappa}{\omega_{1 \vec q}} \delta(\omega - \omega_{1 \vec q})  
- \frac{\gamma_1\!-\!\kappa}{\omega_{2 \vec q}} 
\delta(\omega - \omega_{2 \vec q})\!\biggr]\!.   
\nonumber\\    
\nonumber
\end{eqnarray}   

\section{Summary and discussion} 
\label{sec:discussion} 
In this paper we investigated a spin-orbital superexchange Hamiltonian  
in a Mott insulator with $t_{2g}^1$ electron configuration,  
focusing mainly on the orbital order and dynamics in the spin 
ferromagnetic state. An important feature of the Hamiltonian 
in the spin polarized state is the large frustration of orbital states, 
thus the ground state is governed by the interplay between orbital 
frustration and quantum fluctuations. 
On the classical level, there is a {\it local} $Z_2$ symmetry 
which leads to an infinite degeneracy of classical configurations. 
Long-range orbital order
does occur in the model by a quantum order-from-disorder mechanism, 
which selects a particular ordering patterns. 
Orbital orderings are quite unusual having
highly noncollinear four sublattice structure, and provide 
the same spin couplings in all cubic directions.  

Besides classical local $Z_2$ symmetry which is removed by quantum dynamics,
there are exact conservation laws in the orbital Hamiltonian. They
are related to the conservation of orbital quantum numbers in the SE process, 
and lead to a multitude of degenerate quantum ground states which can 
smoothly be connected to each other by phase rotations of the complex 
orbital order parameter. Such continuous rotations generate orbital
Goldstone modes, which have 2D dispersion because of planar geometry of 
$t_{2g}$ orbitals. As a result, static orbital order sets in at zero
temperature only. Degenerate quantum ground states are physically different: 
depending on the phase of the orbital condensate, they describe 
quadrupole or magnetic orderings or their coherent mixture. Extrinsic
perturbations, e.g. lattice distortions or spin-orbit interactions
may remove the degeneracy and fix the phase of the condensate.
Reflecting the large quantum fluctuations,  
the orbital order parameter is unusually small. 

We found that the   
orbitally ordered ferromagnetic state is slightly higher in energy than 
the spin-AF orbital liquid state. This is because the latter state gains
an additional quantum energy from coupled spin-orbital fluctuations.
To explain ferromagnetism of YTiO$_3$, we emphasized  
the role played by Ti-O-Ti bond angle distortion. This distortion 
favors ferrostate by generating an unfrustrated ferromagnetic SE interaction 
via virtual hopping of electrons between NN $t_{2g}$ and $e_g$ orbitals.
Even more importantly, the bond distortion eliminates orbital soft modes and 
opens a large orbital gap, such that orbital order becomes stable 
at finite temperature. This distortion stabilizes the quadrupole ordered state.

\begin{figure} 
\epsfxsize=0.6\columnwidth 
\centerline{\epsffile{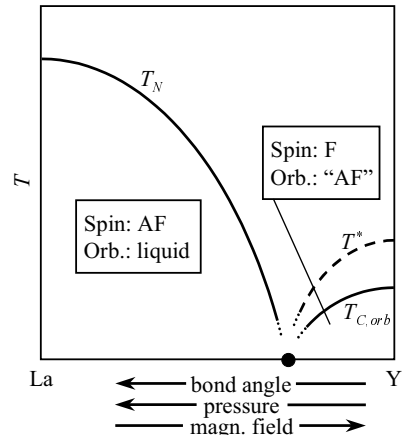}} 
\caption{Proposed picture for the evolution of magnetic and orbital states 
in perovskite titanates.  
$T_C$ and $T_N$ are Curie and N{\' e}el spin ordering temperatures,  
respectively. Orbital ordering below $T_{orb} \simeq T_C$ is expected 
in ferromagnetic region. Below $T^*$, short range ferro-type spin 
and noncollinear, dominantly ``AF''-type orbital correlations grow up. }  
\label{fig:temp} 
\end{figure}

The strong competition between AF to F states in the present model 
has direct relevance to nearly continuous transition between these states 
observed in titanates. In these compounds, 
$A$-site substitution from La to Y increases  
the Ti-O-Ti bond distortion, hence changing gradually a delicate balance
between AF and F states. Because of the orbital fluctuations, 
spin-exchange integral on every link experiences strong fluctuations, both 
in amplitude and in sign, and the system may develop either an AF or F state
depending on local orbital correlations.  
In this picture of ``fluctuating exchange bonds,'' the 
magnetic transition temperatures, $T_N$ and $T_C$, represent only a time-averaged 
static component of the spin couplings. 
Its value is only a fraction 
of full superexchange energy scale, and can gradually be
tuned by external forces such as lattice distortion, pressure, {\it etc}. 
We think that weak orbital order may continuously evolve in titanates 
when the bond angle decreases below a certain critical value, 
and propose the phase diagram shown in Fig.~\ref{fig:temp}. 
The sign of the time-averaged spin coupling depends on a local correlation 
of orbitals. To the right of the critical point orbital correlations 
are more antiferromagnetic, supported by noncollinear orbital orderings.  
To the left, the genuine ground state of the $t_{2g}$ superexchange,  
an orbital disordered state supporting spin AF is stabilized.  
In the proximity area, a fluctuating part of the overall superexchange 
interaction dominates, and separation of the spin and orbital degrees of 
freedom might no longer be possible.  
This scenario can be tested experimentally 
by investigating the spin and orbital transition temperatures 
under high pressure and magnetic field. On the theoretical side, 
a quantitative description of the transition between ferromagnetic and AF states, 
controlled by orbital order-disorder transition, remains an
interesting and challenging problem. 

Further, in order to discuss the recent spin-wave data in YTiO$_3$, 
we derived the low-energy spin Hamiltonian by including a relativistic 
spin-orbit coupling and lattice distortions that induce the orbiton gap.
Using this Hamiltonian, we calculated the magnon gap and found that it is 
very small because of high symmetry of the underlying orbital order.  
Both the real and the spin space isotropy of the spin-wave spectra observed 
experimentally find natural explanations within the proposed theory. 

We also calculated the orbital contribution to resonant x-ray intensity
and to dynamical magnetic structure factor. Predictions made should be helpful
in further experimental study of titanates.

\acknowledgments 
We would like to thank B.~Keimer and S. Maekawa for stimulating discussions.
Discussions with C.~Ulrich, S.~Ishihara, T.~Kiyama, M.~Itoh and J.~Akimitsu  
are also acknowledged.  
One of us (G.~Kh.) would like to thank International Frontier Center for 
Advanced Materials at the Institute for Materials Research, 
Tohoku University, where a part of this work was carried out, 
for its kind hospitality.  
S.~O. acknowledges the hospitality of Max-Planck-Institut  
f\"ur Festk\"orperforschung in Stuttgart during his several visits. 
 
\appendix 

\section{Bond dependence of quadrupole moment operators} 
\label{sec:QTgamma} 
In this appendix,  
we present the explicit expressions for $Q_\alpha^{(\gamma)}$, $T_\alpha^{(\gamma)}$.  
By replacing $l_\alpha$ in Eqs.~(\ref{eq:quadru}) with  
$l_\alpha^{(\gamma)}$, one obtains  
\begin{eqnarray} 
Q_x^{(\gamma)}= \left\{ \begin{array}{rl} 
-c Q_x \mp s T_z, & \quad \mbox{for $\gamma = a (b)$} \\ 
Q_x, & \quad \mbox{for $\gamma = c$}  
\end{array} \right. \nonumber \\ 
T_z^{(\gamma)}= \left\{ \begin{array}{rl} 
-c T_z \pm s Q_x, & \quad \mbox{for $\gamma = a (b)$} \\ 
T_z, & \quad \mbox{for $\gamma = c$}  
\end{array} \right. \nonumber \\ 
T_x^{(\gamma)}= \left\{ \begin{array}{rl} 
-c T_x \mp s T_y, & \quad \mbox{for $\gamma = a (b)$} \\ 
T_x, & \quad \mbox{for $\gamma = c$}  
\end{array} \right. \nonumber \\ 
T_y^{(\gamma)}= \left\{ \begin{array}{rl} 
-c T_y \pm s T_x, & \quad \mbox{for $\gamma = a (b)$} \\ 
T_y. & \quad \mbox{for $\gamma = c$}  
\end{array} \right.  
\end{eqnarray} 
$T_{\pm 1}^{(\gamma)}$ is given by  
$T_{\pm 1}^{(\gamma)} = T_y^{(\gamma)} \pm T_x^{(\gamma)}$.  
 
\section{Coefficients of Bogoliubov transformation} 
\label{sec:Bogoliubov} 
We present the coefficients of Bogoliubov transformation which diagonalize 
$H_{OW}$ in Eq.~(\ref{eq:How2}). This transformation reads as follows:
\begin{eqnarray} 
\tilde a_{\vec k} &=&  
 u \cosh \theta_1 \alpha_{1 \vec k} + v \cosh \theta_2 \alpha_{2 \vec k} 
-u \sinh \theta_1 \alpha_{1 \: -\vec k}^\dag \nonumber \\ 
&& - v \sinh \theta_2 \alpha_{2 \: -\vec k}^\dag \, , \nonumber \\ 
\tilde b_{\vec k} &=& 
-v \cosh \theta_1 \alpha_{1 \vec k} + u \cosh \theta_2 \alpha_{2 \vec k} 
+v \sinh \theta_1 \alpha_{1 \: -\vec k}^\dag \nonumber \\ 
&& - u \sinh \theta_2 \alpha_{2 \: -\vec k}^\dag \, .  
\end{eqnarray} 
The inverse transformation is given as follows:  
\begin{eqnarray} 
\alpha_{1 \vec k} &=&  
 u \cosh \theta_1\tilde a_{\vec k} - v \cosh \theta_1\tilde b_{\vec k} 
+u \sinh \theta_1\tilde a_{-\vec k}^\dag \nonumber \\ 
&& - v \sinh \theta_1\tilde b_{-\vec k}^\dag \, , \nonumber \\ 
\alpha_{2 \vec k} &=&  
 v \cosh \theta_2\tilde a_{\vec k} + u \cosh \theta_2\tilde b_{\vec k} 
+v \sinh \theta_2\tilde a_{-\vec k}^\dag \nonumber \\ 
&& + u \sinh \theta_2\tilde b_{-\vec k}^\dag \, .  
\end{eqnarray} 
Here, $u$ and $v$ are  
\begin{eqnarray} 
u = {1 \over \sqrt{2}} \sqrt{1 + \frac{\gamma_2}{\kappa}}   
\end{eqnarray} 
and 
\begin{eqnarray} 
v = {1 \over \sqrt{2}} \sqrt{1 - \frac{\gamma_2}{\kappa}}  
sgn(\gamma_3) ,  
\end{eqnarray} 
respectively, and $\theta_{1,2}$ is given by  
$\tanh 2 \theta_{1,2} = \gamma_1 \pm \kappa$.  
 
\section{Magnon gap by single-mode approximation} 
\label{sec:SMA} 
We examine here the magnon excitation gap, generated by  
``cubic'' anisotropy interaction given in Sec.~\ref{subsec:ani},
by using a different approach: namely,  
we apply single-mode approximation.\cite{FEY72} 
In this approximation,  
spin excitation energy at ${\vec q} \rightarrow 0$ is given as follows:  
\begin{eqnarray} 
\omega_{{\vec q} \rightarrow 0} 
= \frac{ \Big\langle \Bigl[ \Bigl[S_{\vec q}^+, H_s \Bigr],  
S_{- \vec q}^- \Bigr] \Big\rangle_{{\vec q} \rightarrow 0} } 
{\Big\langle S_{\vec q}^+ S_{- \vec q}^- \Big\rangle_{{\vec q} \rightarrow 0}} , 
\label{eq:SMA_SW} 
\end{eqnarray} 
where  
$H_s$ is given in Eq.~(\ref{eq:Hs})  
with $S_i^{(a)}=S_i^x, S_i^{(b)}=S_i^y$, and $S_i^{(c)}=S_i^z$.  
The double-commutator correlation function in the numerator of Eq.~({\ref{eq:SMA_SW}})  
is equal to $\sum_{ij} \Gamma_{ij}^{(\gamma)}$, where  
\begin{eqnarray} 
\Gamma_{ij}^{(a)} &=& -\tilde A \Big\langle S_i^z S_j^z -  
{1 \over 2} \Bigl( S_i^- S_j^+ + S_i^- S_j^- \Bigr) \Big\rangle , \nonumber \\ 
\Gamma_{ij}^{(b)} &=& -\tilde A \Big\langle S_i^z S_j^z -  
{1 \over 2} \Bigl( S_i^- S_j^+ - S_i^- S_j^- \Bigr) \Big\rangle , \nonumber \\ 
\Gamma_{ij}^{(c)} &=& -\tilde A \Big\langle -2 S_i^z S_j^z + S_i^+ S_j^- \Big\rangle .  
\end{eqnarray} 
By using linear spin-wave theory,  
one obtains  
\begin{eqnarray} 
\sum_{\langle ij \rangle} \Gamma_{ij}^{(\gamma)} 
&=& 2 z S\tilde A \sum_{\vec k} \biggl[ 
\gamma_3 \Big\langle a_{\vec k}^\dag a_{\vec k} \Big\rangle 
-\frac{1}{\sqrt{3}} \gamma_2\Big\langle a_{\vec k}^\dag a_{- \vec k}^\dag \Big\rangle \biggr] ,  
\nonumber \\ 
\Big\langle S_{\vec q}^+ S_{- \vec q}^- \Big\rangle_{{\vec q} \rightarrow 0} 
&=& 2S \big\langle 1 + a_0^\dag a_0 \big\rangle .  
\end{eqnarray}  
Expectation values $\langle a_{\vec k}^\dag a_{\vec k} \rangle$  
and $\langle a_{\vec k}^\dag a_{- \vec k}^\dag \rangle$ are calculated  at $T=0$.  
Up to linear in $\tilde A$ terms, one obtains $\langle a_{\vec k}^\dag a_{\vec k} \rangle = 0$ and  
\begin{eqnarray}  
\Big\langle a_{\vec k}^\dag a_{- \vec k}^\dag \Big\rangle  
={\tilde A \over 2zJ}{c_x-c_y \over 1 - \gamma_1}.  
\end{eqnarray} 
Consequently, we find the magnon gap
\begin{eqnarray} 
\omega_{{\vec q} \rightarrow 0} = {2\tilde A^2R \over zJ}, 
\end{eqnarray} 
where $R$ is given by Eq.~(\ref{eq:R}).  
This is exactly the result obtained in Sec.~\ref{subsec:spinwave}. 
 
\section{Effect of orbital excitations on the magnon dispersion} 
\label{sec:coherent} 
In Sec.~\ref{subsec:iso}, we discussed the renormalization of 
nearest-neighbor isotropic spin coupling $J$ by orbital fluctuations.  
Here, we investigate this effect in more detail, by considering 
effects of the dynamical spin/orbital coupling on magnon spectra.
In terms of magnon $s_{\vec p}$ and orbiton 
$\tilde a_{\vec q}, \tilde b_{\vec q}$
operators, the dynamical spin/orbital coupling  
in Eq.(\ref{eq:mag_orb}) is expressed as (in units of $r_1 J_{SE}$) 
\begin{eqnarray} 
H_{int} &=&-\frac{1}{2} 
\sum_{\vec p \vec q \vec q'} s_{\vec p'}^\dag s_{\vec p} 
\Bigl[ Q \Gamma_0  
\Bigl(\tilde a_{\vec q'}^\dag \tilde a_{- \vec q} + 
\tilde b_{\vec q'}^\dag \tilde b_{-\vec q} \Bigr) \nonumber \\  
&& \hspace{2.5em} + \bigl( \Gamma_1 + \Gamma_2 \bigr)  
\tilde a_{\vec q'}^\dag \tilde a_{\vec q}^\dag  
+ \bigl( \Gamma_1 - \Gamma_2 \bigr)  
\tilde b_{\vec q'}^\dag \tilde b_{\vec q}^\dag  \nonumber \\ 
&& \hspace{2.5em} - 2 \Gamma_3  
\tilde a_{\vec q'}^\dag \tilde b_{\vec q}^\dag \Bigr]  + H.c.   
\label{eq:Hint} 
\end{eqnarray} 
where $\vec p' = \vec p - \vec q - \vec q'$.  
Factor $Q \simeq$ 0.19 stems from Hartree decoupling of the $Q_{iz}Q_{jz}$ term  
in Eq.~(\ref{eq:Hrot3}).  
$\Gamma_0$ and $\Gamma_{i(=1,2,3)}$ are given as follows:  
\begin{eqnarray} 
\Gamma_0 = 1 + \gamma_{1 \: \vec q+\vec q'} - \gamma_{1 \: \vec p} -  
\gamma_{1 \: \vec q+ \vec q'-\vec p} \nonumber \\ 
\Gamma_{i} =  
\gamma_{i \: \vec q} + \gamma_{i \: \vec q'} - \gamma_{i \: \vec q-\vec p} - \gamma_{i \: \vec q'-\vec p}.  
\end{eqnarray} 
 
Second-order perturbation with respect to $H_{int}$ gives 
the renormalization of magnon excitation energy as  
$\omega_{\vec p} = \omega_{\vec p}^{(0)} -\delta \omega_{\vec p}$.  
Here, magnon softening $\delta \omega_{\vec p}$ is given by  
\begin{eqnarray} 
\delta \omega_{\vec p} &=&  
\frac{1}{2} \sum_{\vec q \vec q'}  
\Biggl[ \frac{M (1+\lambda)}{\omega_{1 \vec q} + \omega_{1 \vec q'} + \omega_{\vec p'} -\omega_{\vec p}}  
\nonumber \\ 
&&\hspace{2.5em}  
+ \frac{\widetilde M (1+\tilde \lambda)} 
{\omega_{1 \vec q} + \omega_{2 \vec q'} + \omega_{\vec p'} -\omega_{\vec p}} \Biggr],   
\label{eq:delta_w} 
\end{eqnarray} 
where $\omega_{1,2 \, \vec q}=(\mu^2-\gamma_{\pm}^2)^{1/2}$ with  
$\gamma_{\pm}=\gamma_1 \pm \kappa$ correspond to two orbiton branches.   
The orbiton chemical potential $\mu$ controls the orbital gap. Magnon energy
$\omega_p$ that enters in this equation is considered to have a
NN Heisenberg form 3$|J|(1-\gamma_{1 \vec q})$.  
The matrix elements in the numerator have a following structure: 
\begin{eqnarray} 
M= (\Gamma_1 + r \Gamma_0)^2 (1+x) + \Gamma_2^2 (1+y) + \Gamma_3^2 (1-y) \nonumber \\ 
+ 2 (\Gamma_1 + r \Gamma_0)(\Gamma_2 c_2 + \Gamma_3 c_3)  
+ 2 \Gamma_2 \Gamma_3 z,   
\end{eqnarray} 
where  
\begin{eqnarray} 
\lambda &=& \frac{\mu^2 + \gamma_{+\vec q} \gamma_{+\vec q'} - \omega_{1 \vec q} \omega_{1 \vec q'}} 
{2 \omega_{1 \vec q} \omega_{1 \vec q'}}, \nonumber \\ 
r &=& - \frac{\mu}{1+\lambda} 
\frac{\gamma_{+\vec q}+\gamma_{+\vec q'}}{2 \omega_{1 \vec q} \omega_{1 \vec q'}} Q,  \nonumber \\ 
x&=&(\gamma_{2 \, \vec q} \gamma_{2 \, \vec q'} + \gamma_{3 \, \vec q} \gamma_{3 \, \vec q'}) 
/ \kappa_{\vec q} \kappa _{\vec q'}, \nonumber\\ 
y&=&(\gamma_{2 \, \vec q} \gamma_{2 \, \vec q'} - \gamma_{3 \, \vec q} \gamma_{3 \, \vec q'}) 
/ \kappa_{\vec q} \kappa _{\vec q'}, \nonumber\\ 
z&=&(\gamma_{2 \, \vec q} \gamma_{3 \, \vec q'} + \gamma_{3 \, \vec q} \gamma_{2 \, \vec q'}) 
/ \kappa_{\vec q} \kappa _{\vec q'},   
\end{eqnarray} 
and  
\begin{eqnarray} 
c_2=\frac{\gamma_{2 \, \vec q}}{\kappa_{\vec q}} + \frac{\gamma_{2 \, \vec q'}}{\kappa_{\vec q'}},  
\nonumber\\ 
c_3=\frac{\gamma_{3 \, \vec q}}{\kappa_{\vec q}} + \frac{\gamma_{3 \, \vec q'}}{\kappa_{\vec q'}}.  
\end{eqnarray} 
 
Interband orbiton transitions are represented by a second term in Eq.~(\ref{eq:delta_w}). 
To obtain $\widetilde M$ and $\tilde \lambda$, one should just replace 
$\kappa_{\vec q'}$ $\rightarrow$ $-\kappa_{\vec q'}$ in above equations. 
(This also leads to $\omega_{1 \, \vec q'}$  
$\rightarrow$ $\omega_{2 \, \vec q'}$ in $\lambda$ and $r$).  
One can verify that the function $\delta \omega_{\vec p}$ has a cubic symmetry 
in a momentum space.  
This property is guaranteed by the high symmetry of the underlying 
orbital order in any level of approximations.  
The longer-range, next-NN spin couplings might, of course,  
be dynamically generated by orbital fluctuations. 
 
\begin{figure} 
\epsfxsize=0.8\columnwidth 
\centerline{\epsffile{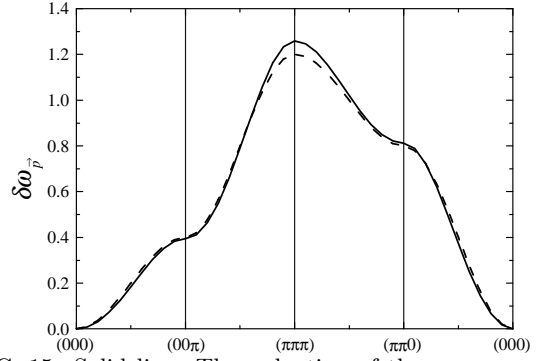}} 
\caption{Solid line: The reduction of the magnon energy $\delta \omega_p$ 
(in units of $r_1 J_{SE}$), calculated from Eq.(\ref{eq:delta_w}).  
Broken line is a function $3(\delta J)(1- \gamma_{1 \, \vec p})$  
with $\delta J=0.2$, showing that the effect of $\delta \omega_p$
can fairly be regarded as an effective reduction of NN spin coupling.}  
\label{fig:delta_w} 
\end{figure}

We show the numerical result for magnon renormalization
$\delta \omega_{\vec p}$ in Fig.~\ref{fig:delta_w}.  
For magnon dispersion in Eq.(\ref{eq:delta_w}) we used 
$|J|=0.03 (r_1 J_{SE}^{(0)}) \simeq 0.05 (r_1J_{SE})$ as obtained 
from Eq.~(\ref{eq:Jtheta}), while orbiton dispersion is calculated
with $\mu =1.41$ which gives orbiton 
gap $\simeq r_1 J_{SE}$. As estimated in the main text, 
such a gap would be induced by nondiagonal hopping
To see a deviation of the magnon remormalization $\delta \omega$  
from the NN Heisenberg form, 
broken line shows a function $3(\delta J)(1- \gamma_{1 \, \vec q})$  
with $\delta J=0.2$.  
It is noticed that $\delta J$ is indeed close to $\delta J_{eff} = 2/9$  
obtained in the main text for the reduction of NN spin couplings. 
Slight deviations from  
simple NN model are however visible, in particular a stronger  
softening at the $(\pi\pi\pi)$ point is seen.  
By a numerical fitting, these deviations can be traced  
back to the appearance of longer-range ferromagnetic couplings  
$J_2=-0.003$, $J_3$=0, and $J_4=-0.007$ (all in units of $r_1 J_{SE}$).  
This is understood due to longer-range orbital singlet correlations  
along the cubic directions, as $J_4$ corresponds to a ferromagnetic  
coupling between second-nearest-neighbor spins along cubic axes. 
In principle, these corrections could be observable as a slight  
enhancement of the ratio of small momentum spin stiffness 
to full magnon bandwidth as compare to the NN Heisenberg model. 
We should notice however that more quantitative predictions are not possible 
at the present stage of the theory for rather obvious reason: 
A dynamical spin-orbital interaction is strong (with the coupling constant
being of the order of 1), so a more elaborate treatment is needed to 
quantify the strongly correlated model under consideration.   
 
\section{Examination of spin interactions in the previous orbital models 
for YTiO$_3$}  
\label{sec:AniMiz} 
In this appendix, we examine the magnetic interactions  
in the orbital state previously reported  
in Refs.~\onlinecite{MIZ96,SAW97}-\onlinecite{NAK02}, in order to check
whether this state can explain recent neutron-scattering results
on the spin couplings, spin canting, and magnon gap.  
 
{\em \underline{The Heisenberg spin exchange coupling}}: 
First, we discuss bond dependence of the isotropic spin interactions.  
The orbital state reported by Hartree-Fock\cite{MIZ96} and band-structure\cite{SAW97} 
calculations are expressed as  
\begin{eqnarray} 
|\psi_{1,3} \rangle = \sqrt{n_c} | d_{xy} \rangle \pm \sqrt{1-n_c} | d_{xz} \rangle , \nonumber \\ 
|\psi_{2,4} \rangle = \sqrt{n_c} | d_{xy} \rangle \pm \sqrt{1-n_c} | d_{yz} \rangle  
\label{eq:miz_function} 
\end{eqnarray} 
with $n_c$ being an occupation of the $xy$ orbital.  
Using these wave functions, it is easy to obtain from Eq.~(\ref{eq:Jgamma})  
spin exchange couplings along $c$-, $a$ and $b$-axes:  
\begin{eqnarray} 
J^{(c)} = \frac{1}{2} J_{SE}[(r_1+r_2 r_3)(1-n_c) - (r_1-r_2)](1-n_c) 
\end{eqnarray} 
and  
\begin{eqnarray} 
J^{(ab)} = \frac{1}{2} J_{SE} \Bigl[(r_1+r_2 r_3) n_c^2 - \frac{1}{2}(r_1-r_2)(1+n_c) \Bigr]. 
\end{eqnarray} 
 
\begin{figure} 
\epsfxsize=0.8\columnwidth 
\centerline{\epsffile{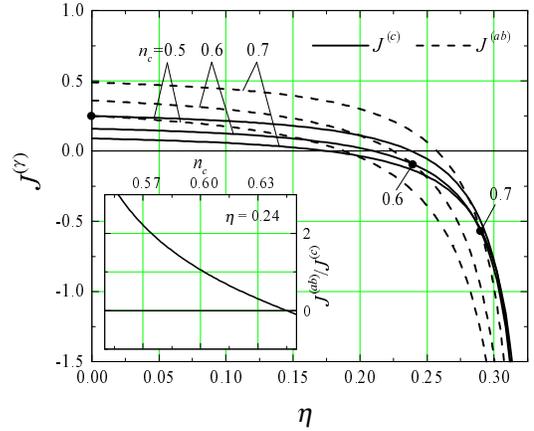}} 
\caption{Spin exchange interactions in the model (\ref{eq:miz_function})
as a function of the Hund's coupling $\eta$ for different values
of the $xy$-orbital occupation $n_c$. Energy unit is $J_{SE}=4t^2/U$. 
Filled circles show the position where spin couplings are isotropic 
for fixed value of $n_c$.  
Inset: The ratio $J^{(ab)} /J^{(c)}$ as a function of $n_c$  
for fixed $\eta=0.24$.}  
\label{fig:exchange} 
\end{figure}

The exchange interactions are presented as functions of $\eta$ in 
Fig.~\ref{fig:exchange} for  
different values of $n_c$.  
The ``meeting'' points, where $J^{(c)} = J^{(ab)}$, are shown by
circles for each $n_c$.  
One finds that the isotropy point for the state with $n_c=0.5$ 
(suggested in Refs.~\onlinecite{MIZ96,SAW97}) is right at the border $\eta=0$, 
but the exchange coupling is of the AF sign there.  
For larger $n_c$ one may obtain the isotropy point with F coupling,  
but this requires too large values of $\eta$. Moreover, the ``meeting'' point 
is extremely sensitive to both $\eta$ and $n_c$ and requires fine tuning. 
As shown    
in the inset of Fig.~\ref{fig:exchange},  
the $J^{(ab)} /J^{(c)}$ ratio drastically changes even at a small (just within
$\pm 5$\%) variation of $n_c$, and may even reverse the sign. 
 
{\em \underline{Spin anisotropy interactions}}:  
Next, we investigate the effects of spin anisotropy interactions 
in the state, Eq.~(\ref{eq:miz_function}).
We denote the occupied orbital state described by wave functions in 
Eq.~(\ref{eq:miz_function}) by $\alpha$, while the lowest unoccupied
state is called $\beta$. The state $\beta$ has wave function which is a  
counterpart of that for $\alpha$ state; e.g., 
$|\tilde{\psi}_1\rangle =\sqrt{1-n_c}|d_{xy}\rangle -\sqrt{n_c}|d_{xz}\rangle$ 
on site 1.  
Level separation between $\alpha$ and $\beta$ is introduced 
as $\Delta_{\alpha \beta}$.  
Active components of the angular momenta at sites 1, 3 and 2, 4 
are $l_x$ and $l_y$, respectively.  
These are expressed in terms of orbital doublet operators as follows:  
\begin{equation} 
(l_x)_{1,3} = \pm i (\beta^\dag \alpha - \alpha^\dag \beta) ,  
\label{eq:l13} 
\end{equation} 
and  
\begin{equation} 
(l_y)_{2,4} = \mp i (\beta^\dag \alpha - \alpha^\dag \beta) . 
\label{eq:l24} 
\end{equation} 
As usual, the spin anisotropy Hamiltonian follows from perturbation theory  
with respect to spin-orbit coupling $H_{so}$ and superexchange interaction 
$H_{SE}$.  
In the SE operators $\hat J_{ij}^{(c)}$ and $\hat K_{ij}^{(c)}$,  
the following terms contribute to the spin anisotropy Hamiltonian:  
(i) the terms operating in the $\beta$ excited states, and  
(ii) the terms connecting a ground state $\alpha$ with excited ones $\beta$.   
 
{\em Symmetric spin anisotropy interaction}:  
Third-order perturbation with respect to $H_{so}$ and $H_{SE}$  
gives symmetric spin anisotropy Hamiltonian $H_{ani}$.  
For 1-3 bond along the $c$-direction, $H_{ani}$ is given by  
\begin{eqnarray} 
H_{ani}^{(13)} =  
\widetilde A \bigl( S_{1x} S_{3x} \bigr) , 
\end{eqnarray} 
where  
\begin{equation} 
\widetilde A =  J_{SE} \frac{r_1+r_2}{4} 
\Bigl(\frac {\lambda}{\Delta_{\alpha \beta}}\Bigr)^2 4 n_c (1-n_c).  
\label{Amiz}
\end{equation} 
Here, the terms proportional to small numbers $(r_1-r_2)$  
and $(r_2-r_3)$ are neglected.  
For 2-4 bond along $c$-direction  
\begin{eqnarray} 
H_{ani}^{(24)} =  
\widetilde A \bigl( S_{2y} S_{4y} \bigr).  
\end{eqnarray} 
Note that spin components correspond to that of active angular momentum  
and the interaction is of the AF sign.  
This originates from the fluctuation of active angular momenta with 
AF correlation between NN sites, as can be inferred from  
Eqs.~(\ref{eq:l13}) and (\ref{eq:l24}).
For the 1-2 bond in the $ab$-plane, $H_{ani}$
has the following form:  
\begin{eqnarray} 
H_{ani}^{(12)} =  
- \frac {1}{4}\widetilde A \bigl( S_{1z} S_{2z} \bigr) 
+ \frac{1}{2} \widetilde A \bigl( S_{1y} S_{2x} \bigr) . 
\end{eqnarray} 
The combination of spin components in the last term is different from that  
of active momenta at sites 1 and 2.  
Such terms originate from the following processes (combination of operators):  
$\langle l_{1x} S_{1x} (S_{1z} S_{2z}) l_{2y} S_{2y} \rangle_{orb}$,  
$\langle l_{2y} S_{2y} (S_{1z} S_{2z}) l_{1x} S_{1x} \rangle_{orb}$,  
{\it etc}.  
For 3-4 bond, symmetric spin anisotropy Hamiltonian $H_{ani}$ has the same 
form as $H_{ani}^{(12)}$,  
where site index 1 (2) is replaced with 3 (4).  
It can be shown that $H_{ani}$ does not cause finite spin canting 
when $|J| \agt \widetilde A/2$, where $J$ is the isotropic spin coupling.  
However, it leads to a large magnon gap of classical origin, which we
obtained to be $\sqrt{3}\widetilde AS$.

{\em DM interaction}:  
For the 1-3 bond, second-order perturbation with respect to $H_{so}$  
and $H_{SE}$ gives  
\begin{eqnarray} 
H_{DM}^{(13)} =  
\widetilde D^{(c)} \vec d_{13} \cdot \Bigl( \vec S_1 \times \vec S_3 \Bigr), 
\label{eq:DM13} 
\end{eqnarray} 
where constant $\widetilde D^{(c)}$ is obtained as 
\begin{equation} 
\widetilde D^{(c)} = J_{SE} \frac{r_1+r_2}{4} 
\frac{\lambda}{\Delta_{\alpha \beta}}  
4n_c^{1/2}(1-n_c)^{3/2}.  
\label{Dmiz}
\end{equation} 
DM vector $\vec d_{13}$ is given as $\vec d_{13}=(-1,0,0)$.  
We point out here that DM constant $\widetilde D^{(c)}$ is much 
larger than $D$ in Eq.~(\ref{eq:defD}), obtained in the main text 
for the SE-driven orbital states. The reason is that the orbital order
given by Eq.(\ref{eq:miz_function}) has not that high symmetry, and
the terms proportional to $(r_1+r_2)$ in the operator $\hat J_{ij}^{(c)}$ do
contribute to DM interaction. Therefore, the ratio 
$\widetilde D^{(c)}/D\propto \frac{r_1+r_2}{r_1-r_2}\propto 1/\eta$ is large.  

\begin{figure} 
\epsfxsize=0.6\columnwidth 
\centerline{\epsffile{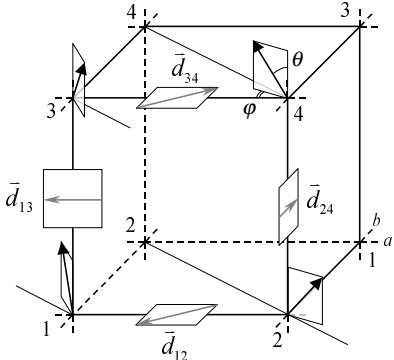}} 
\caption{Antisymmetric DM interaction pattern obtained for the orbital model
(\ref{eq:miz_function}).
Gray arrow denoted by $\vec d_{ij}$ shows the orientation of DM vectors.
Black arrows represent the direction of spins favorable 
for these interactions.}  
\label{fig:AniMiz} 
\end{figure}

On the in-plane bond 1-2, we find
\begin{eqnarray} 
H_{DM}^{(12)} =  
\frac{1}{2} \widetilde D^{(ab)} \vec d_{12} \cdot \Bigl( \vec S_1 \times \vec S_2 \Bigr), 
\label{eq:DM12} 
\end{eqnarray} 
with  
$\widetilde D^{(ab)} = \widetilde D^{(c)} n_c/(1-n_c)$ 
and $\vec d_{12}=(-1,-1,0)$.  
The DM interactions on the 
2-4 and 3-4 bonds are given by the same forms as  
Eqs.~(\ref{eq:DM13}) and (\ref{eq:DM12}), respectively,   
where $\vec d_{24} = (0,1,0)$ and $\vec d_{34} = - \vec d_{12}$.  
The DM interaction and related spin structure in the orbital model  
(\ref{eq:miz_function}) are schematically shown in Fig.~\ref{fig:AniMiz}.  

{\em \underline{Spin canting and magnon gap}}: 
First we estimate anisotropy constants. We consider the orbital state 
with $n_c=0.6$ which has a chance to explain the $J^{(c)} = J^{(ab)}$ property,
giving isotropic spin coupling $J$ about $-0.1J_{SE}$ at $\eta =0.24$ 
(see Fig.~\ref{fig:AniMiz}). Given these parameters, one obtains from 
Eqs.~(\ref{Amiz}) and (\ref{Dmiz}),
\begin{eqnarray} 
\widetilde A &\simeq& 11 (\lambda/\Delta_{\alpha\beta})^2 |J|, \nonumber \\
\widetilde D^{(c)}&\simeq& 10 (\lambda/\Delta_{\alpha\beta}) |J|, \nonumber \\
\widetilde D^{(ab)}&=&(3/2)\widetilde D^{(c)}.
\end{eqnarray} 

By minimizing classical energy of DM 
and isotropic Heisenberg interactions, 
we find that spins cant away from the $c$ axis by an angle
$\theta \sim \frac{\widetilde D^{(ab)}}{3\sqrt{2}J} 
\sim 3.5(\lambda/\Delta_{\alpha\beta})$. Also, within a 
linear spin-wave theory we estimate a magnon gap generated by symmetric
anisotropy and DM interactions as 
$\sqrt{3} \widetilde AS \sim 9.5 (\lambda/\Delta_{\alpha\beta})^2|J|$ and 
$4S|J|\theta^2 \sim 25 (\lambda/\Delta_{\alpha\beta})^2|J|$, correspondingly.
With $\lambda \simeq 19$~meV and $|J|\simeq 2.75$~meV, we obtain that
the ratio $(\lambda/\Delta_{\alpha\beta})$ must be less than 0.05-0.06 
in order to be consistent with the observed canting angle ($\sim 0.17$~rad) 
and an upper limit for the gap ($\sim 0.3$ meV). 
Thus, the splitting of the lowest orbital 
doublet $\Delta_{\alpha\beta}$ should be at least about 300 meV.
However, this is hard to reconcile with almost equal four short Ti-O bonds
in titanates suggesting an almost degenerate doublet picture.

Based on the above analysis, we think that 
the orbital state (\ref{eq:miz_function}) predicted by band-structure
calculations is not supported
by recent neutron-scattering experiments in YTiO$_3$.\cite{ULR02} 
This is perhaps not really surprising, as a Mott insulator
with orbital degeneracy represents a strongly correlated system, which
is difficult to address in a framework of weakly interacting electrons.


\begin{references}
\bibitem{PAIR} 
In addition to the result of Ref.~\onlinecite{KHA01}, the operator  
$B_{ij}^{(\gamma)}$ in Eq.~(\ref{eq:B_ab}) 
contains more (last two) terms, which  
result from spin-singlet electron pair dynamics 
in the virtual $d^2$ excited state.  
It is noticed that such pair-hopping terms 
vanish in a spin polarized ferromagnetic state.
\end{references}
\end{document}